\documentclass[12pt,a4paper]{article}
\usepackage{amsmath,amsfonts,amssymb,epsfig,subfigure}
\usepackage[printfigures,figmark]{figcaps}

\usepackage{pstricks}

\renewcommand{\vec}[1]{\boldsymbol{#1}}

\newcommand{\imi}{\textrm{i}}

\def \curl{\mbox{curl\hskip 1pt}}
\def \Curl{\mbox{Curl\hskip 1pt}}
\def \div{\mbox{div\hskip 1pt}}
\def \Div{\mbox{Div\hskip 1pt}}
\def \grad{\mbox{grad\hskip 1pt}}
\def \Grad{\mbox{Grad\hskip 1pt}}

\begin{document}

\numberwithin{equation}{section}

\title{Incremental magnetoelastic deformations, \\ with application to surface instability}
\author{
   M.~Ott\'enio, M.~Destrade, R.W.~Ogden,
}
  \date{2007}
\maketitle

\begin{abstract}

In this paper the equations governing the deformations of
infinitesimal (incremental) disturbances superimposed on finite
static deformation fields involving magnetic and elastic
interactions are presented. The coupling between the equations of
mechanical equilibrium and Maxwell's equations complicates the
incremental formulation and particular attention is therefore paid
to the derivation of the incremental equations, of the tensors of
magnetoelastic moduli and of the incremental boundary conditions
at a magnetoelastic/vacuum interface.

The problem of surface stability for a solid half-space under
plane strain with a magnetic field normal to its surface is used
to illustrate the general results. The analysis involved leads to
the simultaneous resolution of a bicubic and vanishing of a
$7\times 7$ determinant.  In order to provide specific
demonstration of the effect of the magnetic field, the material
model is specialized to that of a ``magnetoelastic Mooney-Rivlin
solid''. Depending on the magnitudes of the magnetic field and the
magnetoelastic coupling parameters, this shows that the half-space
may become either more stable or less stable than in the absence
of a magnetic field.

\end{abstract}

\section{Introduction}

One of the main reasons for industrial interest in rubber-like
materials resides in their ability to dampen vibrations and to
absorb shocks. This paper is concerned with an extension of the
nonlinear elasticity theory adopted for describing the properties
of these materials to incorporate nonlinear magnetoelastic effects
so as to embrace a class of solids referred to as
magneto-sensitive (MS) elastomers.  These ``smart" elastomers
typically consist of an elastomeric matrix (rubber, silicon, for
example) with a distribution of ferrous particles (with a diameter
of the order of 1--5 micrometers) within their bulk. They are
sensitive to magnetic fields in that they can deform significantly
under the action of magnetic fields alone without mechanical
loading, a phenomenon known as \textit{magnetostriction}. As a
result, their mechanical damping abilities can be controlled by
applying suitable magnetic fields. This coupling between
elasticity and magnetism was probably first observed by Joule in
1847 when he noticed that a sample of iron changed its length when
magnetized.

In general, the physical properties of magnetoelastic materials
depend on factors such as the choice of magnetizable particles,
their volume fraction within the bulk, the choice of the matrix
material, the chemical processes of curing, etc.; see
\cite{BeBo02} for details, and also \cite{RiJi83} for an
experimental study on a magneto-sensitive elastomer.

The coupling between magnetism and nonlinear elasticity has
generated much interest over the last $50$ years or so, as
illustrated by the works of Truesdell and Toupin \cite{TrTo60},
Brown \cite{Brow66}, Yu and Tang \cite{YuTa66}, Maugin
\cite{Maug88}, Eringen and Maugin \cite{ErMa90}, Kovetz
\cite{Kove00}, and others. The corresponding engineering
applications are more recent (see Jolly \textit{et al.}
\cite{JoCa96}, or Dapino \cite{Dapi02}, for instance) and have
generated renewed impetus in theoretical modelling (see, for
example, Dorfmann and Brigadnov \cite{DoBr04}; Dorfmann and Ogden
\cite{DoOg04}); Kankanala and Triantafyllidis \cite{Kank04}. Here,
we derive the (linearized) equations governing incremental effects
in a magnetoelastic solid subject to finite deformation in the
presence of a magnetic field. These equations are then used to
examine the problem of surface stability of a homogeneously
pre-strained half-space subject to a magnetic field normal to its
(plane) boundary. Related works on this subject include the
studies of McCarthy \cite{McCa67}, van de Ven \cite{dVen75},
Boulanger \cite{Boul77, Boul78}, Maugin \cite{Maug81}, Carroll and
McCarthy \cite{CaMc89} and Das \textit{et al.} \cite{DaAc94}.

We adopt the formulation of Dorfmann and Ogden \cite{DoOg04} as
the starting point for the derivation of the incremental
equations.  This involves a \textit{total stress tensor} and a
\textit{modified strain energy function} or \textit{total energy
function}, which enable the constitutive law for the stress to be
written in a form very similar to that in standard nonlinear
elasticity theory.  The coupled governing equations then have a
simple structure. We summarize these equations in Section
\ref{chap4_secresume}. For incompressible isotropic magnetoelastic
materials the energy density is a function of five invariants,
which we denote here by $I_1$ and $I_2$, the first two principal
invariants of the Cauchy-Green deformation tensors, and $I_4$,
$I_5$, $I_6$, three invariants involving a Cauchy-Green tensor and
the magnetic induction vector. This formulation is similar in
structure to that associated with transversely isotropic elastic
solids (see Spencer \cite{Spen71}). The general incremental
equations of nonlinear magnetoelasticity are then derived in
Section \ref{chap4_inctheo}. Therein we define the various
magnetoelastic `moduli' tensors and provide general incremental
boundary conditions. Care is needed in deriving the boundary
equations since the Lagrangian fields in the solid and the
Eulerian fields in the vacuum must be reconciled.

Section \ref{homog} provides a brief summary of the basic
equations associated with the pure homogeneous plane strain of a
half-space of magnetoelastic material with a magnetic field normal
to its boundary.  In Section \ref{section_surf_stab}, the general
incremental equations are applied to the analysis of surface
stability. Not surprisingly, the resulting bifurcation criterion
is a complicated equation, even when the pre-stress corresponds to
plane strain and the magnetic induction vector is aligned with a
principal direction of strain, as is the case here. The
bifurcation equation comes from the vanishing of the determinant
of a $7 \times 7$ matrix, which must be solved simultaneously with
a bicubic equation. To present a tractable example, we therefore
focus on a ``Mooney-Rivlin magnetoelastic solid" for which the
total energy function is linear in the invariants $I_1$, $I_2$,
$I_4$, and $I_5$. Of course, these invariants are nonlinear in the
deformation and the theory remains highly nonlinear. The bicubic
then factorizes and a complete analytical resolution follows.
In addition to the two elastic Mooney-Rivlin parameters
(material constants), the material model involves two
magnetoelastic coupling parameters.  The stability behaviour of
the half-space depends crucially on the values of these coupling
parameters and also on the magnitude of the magnetic field. In
particular, a judicious choice of parameters can stabilize the
half-space relative to the situation in the absence of a magnetic
field. Equally, the half-space can become de-stabilized for
different choices of the parameters.  Thus, even this very simple
model illustrates the possible complicated nature of the
magnetoelastic coupling in the nonlinear regime.

\section{The equations of nonlinear magnetoelasticity} \label{chap4_secresume}

In this section the equations for nonlinear magnetoelastic
deformations, as developed by Dorfmann and Ogden \cite{DoOg04,
DoOg03a, DoOg03b, OgDo05}, are summarized for subsequent use in
the derivation of the incremental equations.

We consider a magnetoelastic body in an undeformed configuration
$\mathcal{B}_0$, with boundary $\partial \mathcal{B}_0$.  A
material point within the body in that confi\-guration is
identified by its position vector $\vec{X}$. By the combined
action of applied mechanical loads and magnetic fields, the
material is then deformed from $\mathcal{B}_0$ to the
configuration $\mathcal{B}$, with boundary $\partial \mathcal{B}$,
so that the particle located at $\vec{X}$ in $\mathcal{B}_0$ now
occupies the position $\vec{x}=\vec{\chi}(\vec{X})$ in the
\textit{deformed configuration} $\mathcal{B}$. The function
$\vec{\chi}$ describes the static deformation of the body and is a
one-to-one, orientation-preserving mapping with suitable
regularity properties.  The deformation gradient tensor $\vec{F}$
relative to $\mathcal{B}_0$ is defined by $\vec{F}=\text{Grad}
\vec{\chi}$, $F_{i \alpha}=\partial x_i / \partial X_\alpha$,
$\Grad$ being the gradient operator in $\mathcal{B}_0$. The
magnetic field vector in $\mathcal{B}$ is denoted $\vec{H}$, the
associated magnetic induction vector by $\vec{B}$ and the
magnetization vector by $\vec{M}$.

To avoid a conflict of standard notations, the Cauchy-Green
tensors are represented here by lower case characters; thus, the
left and right Cauchy-Green tensors are 
$\vec{b}=\vec{FF}^t$ and $\vec{c}=\vec{F}^t\vec{F}$,
respectively, where $^t$ denotes the transpose. The
Jacobian of the deformation gradient is $J=\det \vec{F}$, and the
usual convention $J>0$ is adopted.

\subsection {Mechanical equilibrium}

Conservation of the mass for the material is here expressed as
\begin{equation}
J \rho = \rho_0,
\end{equation}
where $\rho_0$ and $\rho$ are the mass densities in the
configurations $\mathcal{B}_0$ and $\mathcal{B}$, respectively.
For an incompressible material, $J=1$ is enforced so that
$\rho=\rho_0$.

The equilibrium equation in the absence of mechanical body forces,
is given in Eulerian form by
\begin{equation}
\div \vec{\tau}=\vec{0},\label{divtau}
\end{equation}
where $\vec{\tau}$ is the \textit{total Cauchy stress tensor},
which is \textit{symmetric}, and $\div$ is the divergence operator
in $\mathcal{B}$. The \textit{total nominal stress tensor}
$\vec{T}$ is then defined by
\begin{equation} \label{tensor_T}
\vec{T}=J \vec{F}^{-1} \vec{\tau},
\end{equation}
so that the Lagrangian counterpart of the equilibrium equation
\eqref{divtau} is
\begin{equation}
\Div\vec{T}=\vec{0},\label{DivT}
\end{equation}
$\Div$ being the divergence operator in $\mathcal{B}_0$.

Let $\vec{N}$ denote the unit outward normal vector to
$\partial\mathcal{B}_0$ and $\vec{n}$ the corresponding unit
normal to $\partial\mathcal{B}$.  These are related by Nanson's
formula $\vec{n}{\rm d}a=J\vec{F}^{-t}\vec{N}{\rm d}A$,
where ${\rm d}A$ and ${\rm d}a$ are the associated area elements.
The traction on the area element in $\partial\mathcal{B}$ may be
written $\vec{\tau}\vec{n}{\rm d}a$ or as
$\vec{T}^t\vec{N}{\rm d}A$.  A traction boundary
condition might therefore be expressed in the form
\begin{equation}
\vec{T}^t \vec{N}=\vec{t}_{\text a},\label{T-bc}
\end{equation}
where $\vec{t}_{\text a}$ is the applied traction per unit
reference area.  If this is independent of the deformation then
the traction is said to be a \emph{dead load}.

\subsection{Magnetic balance laws}

In the Eulerian description, Maxwell's equations in the absence of
time dependence, free charges and free currents reduce to
\begin{equation}
\div\vec{B}=0,\quad\curl\vec{H}= \vec{0},\label{maxwell}
\end{equation}
which hold both inside and outside a magnetic material, where
$\curl$ relates to $\mathcal{B}$. Thus, $\vec{B}$ and $\vec{H}$
can be regarded as fundamental field variables.  A third vector
field, the magnetization, when required, can be defined by the
standard relation
\begin{equation}
\vec{B}=\mu_0(\vec{H}+\vec{M}).\label{B_HM}
\end{equation}
We shall not need to make explicit use of the magnetization in
this paper.

Associated with the equations \eqref{maxwell} are the boundary
continuity conditions
\begin{equation}
(\vec{B}-\vec{B}^\star)\cdot\vec{n}=0,\quad
(\vec{H}-\vec{H}^\star)\times\vec{n}=\vec{0}
,\label{BH-bcs-eulerian}
\end{equation}
wherein $\vec{B}$ and $\vec{H}$ are the fields in the material and
$\vec{B}^\star$ and $\vec{H}^\star$ the corresponding fields
exterior to the material, but in each case evaluated on the
boundary $\partial\mathcal{B}$.

Lagrangian counterparts of $\vec{B}$ and $\vec{H}$, denoted
$\vec{B}_l$ and $\vec{H}_l$, respectively, are defined by
\begin{equation}
\vec{B}_l=J \vec{F}^{-1} \vec{B},\quad
\vec{H}_l=\vec{F}^t \vec{H}, \label{Hl_Bl}
\end{equation}
and in terms of these quantities equations \eqref{maxwell} become
\begin{equation}
\Div\vec{B}_l=0,\quad \Curl\vec{H}_l=\vec{0},
\end{equation}
where $\Curl$ is the curl operator in $\mathcal{B}_0$. We note in
passing that a Lagrangian counterpart of $\vec{M}$ may also be
defined, one possibility being
$\vec{M}_l=\vec{F}^t\vec{M}$.

The boundary conditions \eqref{BH-bcs-eulerian} can also be
expressed in Lagrangian form, namely
\begin{equation}
(\vec{B}_l-J\vec{F}^{-1}\vec{B}^\star)\cdot\vec{N}=0,\quad
(\vec{H}_l-\vec{F}^t \vec{H}^\star)\times\vec{N}=\vec{0}
,\label{BH-bcs-lagrangian}
\end{equation}
evaluated on the boundary $\partial\mathcal{B}_0$.

\subsection{Constitutive equations}

There are many possible ways to formulate constitutive laws for
magnetoelastic materials based on different choices of the
independent magnetic variable and the form of energy function. For
present purposes it is convenient to use a formulation involving a
`total energy function', or `modified free energy function', which
is denoted here by $\Omega$, following Dorfmann and Ogden
\cite{DoOg04}. This is defined per unit reference volume and is a
function of $\vec{F}$ and $\vec{B}_l$:
$\Omega(\vec{F},\vec{B_l})$.  This leads to the very simple
expressions
\begin{equation}
\vec{T}=\frac{\partial\Omega}{\partial\vec{F}},\quad
\vec{H}_l=\frac{\partial\Omega}{\partial\vec{B}_l}\label{const1}
\end{equation}
for a magnetoelastic material without internal mechanical
constraints, and
\begin{equation}
\vec{T}=\frac{\partial \Omega}{\partial \vec{F}} - p
\vec{F}^{-1},\quad
\vec{H}_l=\frac{\partial\Omega}{\partial\vec{B}_l}\label{const2}
\end{equation}
for an incompressible material, where $p$ is a Lagrange multiplier
associated with the constraint $\det\vec{F}=1$.  Note that the
expression for $\vec{H}_l$ is unchanged except that now
$\det\vec{F}=1$ in $\Omega$.

The Eulerian counterparts of the above equations are
\begin{equation}
\vec{\tau}=J^{-1}\vec{F} \frac{\partial \Omega}{\partial \vec{F}},
\quad \vec{H}=\vec{F}^{-t} \frac{\partial
\Omega}{\partial \vec{B}_l}\label{const3}
\end{equation}
for an unconstrained material, where
$\vec{F}^{-t}=(\vec{F}^{-1})^t$, and
\begin{equation}
\vec{\tau}=\vec{F} \frac{\partial \Omega}{\partial \vec{F}} - p
\vec{I}, \quad\vec{H}=\vec{F}^{-t} \frac{\partial
\Omega}{\partial \vec{B_l}},\label{const4}
\end{equation}
where $\vec{I}$ is the identity tensor. We emphasize that the
first equation in each of \eqref{const1}--\eqref{const4} has
exactly the same form as for a purely elastic material in the
absence of a magnetic field.

\subsection{Isotropic magnetoelastic materials}

In general the mechanical properties of magnetoelastic elastomers
have features that are similar to those of transversely isotropic
materials. During the curing process a preferred direction is
'frozen in' to the material if the curing is done in the presence
of a magnetic field, which aligns the magnetic particles.  If
cured without a magnetic field then the distribution of particles
is essentially random and the resulting magnetoelastic response is
isotropic.  We focus on the latter case here for simplicity, but
the corresponding analysis for the more general case follows the
same pattern, albeit more complicated algebraically.  A general
constitutive theory for the former situation has been developed by
Bustamante and Ogden \cite{Bust07} and applied to some simple
problems.  For isotropic materials, the energy function $\Omega$
depends only on $\vec{c}$ and $\vec{B}_l \otimes \vec{B}_l$,
through the six invariants
\begin{align}
    &I_1=\text{tr}\, \vec{c} , && I_2=\tfrac{1}{2}\left[(\text{tr}\, \vec{c})^2-(\text{tr}
   \,  \vec{c}^2)\right], &&I_3=\det  \vec{c}=J^2, \notag
   \\[0.1cm]
   &I_4=\vec{B}_l \cdot \vec{B}_l, &&I_5=(\vec{c \, B}_l) \cdot \vec{B}_l,
   &&I_6=(\vec{c}^2 \, \vec{B}_l) \cdot \vec{B}_l.\label{invariants_6}
\end{align}

For incompressible materials, $I_3=1$ and only the five invariants
$I_1$, $I_2$, $I_4$, $I_5$, and $I_6$ remain. The total stress
tensor $\vec{\tau}$ is then expressed as
\begin{equation}
\vec{\tau}=- p \vec{I}+2 \Omega_1 \vec{b} + 2 \Omega_2 ( I_1
\vec{b} - \vec{b}^2)  + 2 \Omega_5 \vec{B} \otimes \vec{B} + 2
\Omega_6 ( \vec{B} \otimes \vec{bB} + \vec{bB} \otimes
\vec{B}),\label{tau_iso}
\end{equation}
where $\Omega_i=\partial \Omega / \partial I_i$, and the total
nominal stress tensor $\vec{T}$ as
\begin{multline}
\vec{T}=- p \vec{F}^{-1}+2 \Omega_1 \vec{F}^t + 2
\Omega_2 ( I_1 \vec{F}^t -\vec{F}^t
\vec{b})
\\+ 2 \Omega_5 \vec{B}_l \otimes \vec{B}  + 2 \Omega_6 ( \vec{B}_l
\otimes \vec{bB} + \vec{F}^t \vec{B} \otimes
\vec{B}).\label{T_iso}
\end{multline}

Finally, the magnetic field vector $\vec{H}$ is found from
\eqref{const4}$_2$ as
\begin{equation} \label{H_iso}
\vec{H}=2(\Omega_4 \vec{b}^{-1} \vec{B} + \Omega_5 \vec{B} +
\Omega_6 \vec{bB}),
\end{equation}
and its Lagrangian counterpart is
\begin{equation}
\vec{H}_l=2(\Omega_4\vec{B}_l+\Omega_5\vec{c}\vec{B}_l+\Omega_6\vec{c}^2\vec{B}_l).
\end{equation}

\subsection{Outside the material}

In vacuum, there is no magnetization and the standard relation
\eqref{B_HM} reduces to
\begin{equation}
\vec{B}^\star=\mu_{0}\vec{H}^\star,\label{BHstar}
\end{equation}
where the star is again used to denote a quantity exterior to the
material. Also, the stress tensor $\vec{\tau}$ is now the
\textit{Maxwell stress} $\vec{\tau}^\star$, given by
\begin{equation}
\vec{\tau}^\star=\mu_{0}^{-1} [ \vec{B}^\star \otimes
\vec{B}^\star - \tfrac{1}{2} (\vec{B}^\star \cdot
\vec{B}^\star)\vec{I} ],\label{taustar}
\end{equation}
which, since $\div\vec{B}^\star=0$ and
$\curl\vec{B}^\star=\vec{0}$, satisfies
$\div\vec{\tau}^*=\vec{0}$.

\section{Incremental equations} \label{chap4_inctheo}

\subsection{Increments within the material}

Suppose now that both the magnetic field and, within the material,
the deformation undergo incremental changes (which are denoted by
superposed dots). Let $\vec{\dot{F}}$ and $\vec{\dot{B}}_l$ be the
increments in the independent variables $\vec{F}$ and $\vec{B}_l$.
It follows from \eqref{const1} that the increment $\vec{\dot{T}}$
in $\vec{T}$ and the increment $\vec{\dot{H}}_l$ in $\vec{H}_l$
are given in the form
\begin{equation}
\vec{\dot{T}}=\vec{\mathcal{A}}\vec{\dot{F}}+\vec{\Gamma}\vec{\dot{B}}_l,\quad
\vec{\dot{H}}_l=\vec{\Gamma}\vec{\dot{F}}+\vec{\mathcal{K}}\vec{\dot{B}}_l,\label{incremental-const}
\end{equation}
where $\vec{\mathcal{A}}$, $\vec{\Gamma}$ and $\vec{\mathcal{K}}$
are, respectively, fourth-, third- and second-order tensors, with
components defined by
\begin{equation}
\mathcal{A}_{\alpha i \beta j}=\frac{\partial^2 \Omega}{\partial
F_{i \alpha} \partial F_{j \beta}}, \quad \Gamma_{\alpha i
\beta}=\frac{\partial ^2 \Omega}{\partial F_{i \alpha} \partial
B_{l_{\beta}}}=\frac{\partial ^2 \Omega}{\partial
B_{l_{\beta}}\partial F_{i \alpha} }, \quad \mathcal{K}_{\alpha
\beta}= \frac{\partial^2 \Omega}{\partial B_{l_{\alpha}} \partial
B_{l_{\beta}}}.\label{moduleAGammaK}
\end{equation}
We refer to these tensors as \textit{magnetoelastic moduli
tensors}. We note the symmetries
\begin{equation}
\mathcal{A}_{\alpha i \beta j}=\mathcal{A}_{\beta j \alpha
i},\quad \mathcal{K}_{\alpha \beta}=\mathcal{K}_{\beta
\alpha},\label{symms}
\end{equation}
and observe that $\vec{\Gamma}$ has no such indicial symmetry. The
products in \eqref{incremental-const} are defined so that, in
component form, we have
\begin{equation} \label{dotTH}
\dot{T}_{\alpha i}=\mathcal{A}_{\alpha i \beta j} \dot{F}_{j
\beta} + \Gamma_{\alpha i \beta} \dot{B}_{l_{\beta}}, \quad
\dot{H}_{l_{\alpha}}=\Gamma_{\beta i\alpha} \dot{F}_{i
\beta}+\mathcal{K}_{\alpha \beta}\dot{B}_{l_{\beta}}.
\end{equation}

For an unconstrained isotropic material, $\Omega$ is a function of
the six invariants $I_{1}, I_{2}, I_{3}, I_{4}, I_{5}, I_{6}$, and
the expressions \eqref{moduleAGammaK} can be expanded in the forms
\begin{align}
&\mathcal{A}_{\alpha i \beta j}=\sum_{m=1,\, m\neq 4}^{6}
\sum_{n=1,\, n\neq 4}^{6}\Omega_{m n} \frac{\partial I_n}{\partial
F_{i \alpha}} \frac{\partial I_m}{\partial F_{j \beta}}+
\sum_{n=1,\, n \neq 4}^{6} \Omega_{n} \frac{\partial^2 I_n}
{\partial F_{i \alpha} \partial F_{j \beta}},\notag
\\[0.1cm]
&\Gamma_{\alpha i \beta}=\sum_{m=4}^{6} \sum_{n=1,\,n\neq 4}^{6}
\Omega_{m n}\frac{\partial I_m}{\partial B_{l_\beta}}
\frac{\partial I_n}{\partial F_{i \alpha}}+ \sum_{n=5}^{6}
\Omega_{n} \frac{\partial^2 I_n}{\partial F_{i \alpha} \partial
B_{l_\beta}},\notag \\[0.1cm]
&\mathcal{K}_{\alpha \beta}=\sum_{m=4}^{6} \sum_{n=4}^{6}
\Omega_{m n}\frac{\partial I_m}{\partial B_{l_\alpha}}
\frac{\partial I_n}{\partial B_{l_\beta}}+ \sum_{n=4}^{6}
\Omega_{n} \frac{\partial^2 I_n}{\partial B_{l_\alpha} \partial
B_{l_\beta}}, \label{AGammaK}
\end{align}
where $\Omega_{n}=\partial \Omega / \partial I_{n}$,
$\Omega_{mn}=\partial^2 \Omega / \partial I_{m} \partial I_{n}$.
Expressions for the first and second derivatives of $I_n,\,
n=1,\dots,6$, are given in the Appendix.

For an \textit{incompressible} material, $\vec{T}$ is given by
\eqref{const2}$_1$ and its increment is then
\begin{equation}
\vec{\dot{T}}=\vec{\mathcal{A}} \vec{\dot{F}} + \vec{\Gamma}
\vec{\dot{B}}_l -\dot{p} \vec{F}^{-1} + p \vec{F}^{-1}
\vec{\dot{F}} \vec{F}^{-1}, \label{dTinc}
\end{equation}
which replaces \eqref{incremental-const} in this case. On the
other hand, $\vec{H}_l$ is still given by \eqref{const1}$_2$ and
its increment is unaffected by the constraint of
incompressibility, except, of course, since $\Omega$ is now
independent of $I_{3}=1$, the summations in equations
\eqref{AGammaK} omit $m=3$ and $n=3$.

It is now a simple matter to obtain the incremental forms of the
(Lagrangian) governing equations. We have
\begin{equation}
\Div \vec{\dot{T}}=\vec{0}, \quad \Div \vec{\dot{B}}_l=0,\quad
\Curl \vec{\dot{H}}_l=\vec{0}.\label{cond3delta}
\end{equation}
These equations can be transformed into their Eulerian
counterparts (indicated by a zero subscript) by means of the
transformations
\begin{equation}
\vec{\dot{T}}_{0}=J^{-1} \vec{F} \vec{\dot{T}}, \quad
\vec{\dot{B}}_{l0}= J^{-1} \vec{F} \vec{\dot{B}}_l, \quad
\vec{\dot{H}}_{l0}=\vec{F}^{-t}
\vec{\dot{H}}_l\label{T0Bl0Hl0dot}
\end{equation}
(with $J=1$ for an incompressible material), leading to
\begin{equation}
\div \vec{\dot{T}}_{0}=\vec{0}, \quad \div
\vec{\dot{B}}_{l0}=0,\quad \curl \vec{\dot{H}}_{l0}=\vec{0}
.\label{cond3delta0}
\end{equation}

Now let $\vec{u}$ denote the incremental displacement vector
$\vec{x}-\vec{X}$.  Then,
$\vec{\dot{F}}=\Grad\vec{u}=(\grad\vec{u})\vec{F}$, where $\grad$
is the gradient operator with respect to $\vec{x}$. We use the
notation $\vec{d}$ for the displacement gradient $\grad\vec{u}$,
in components $d_{ij}=\partial u_i/\partial x_j$. From
\eqref{T0Bl0Hl0dot} and \eqref{incremental-const} we then have
\begin{equation}
\vec{\dot{T}}_0=\vec{\mathcal{A}}_0\vec{d}+\vec{\Gamma}_0\vec{\dot{B}}_{l0},\quad
\vec{\dot{H}}_{l0}=\vec{\Gamma}_0\vec{d}+\vec{\mathcal{K}}_0\vec{\dot{B}}_{l0},\label{dotT0Hl0}
\end{equation}
where, in index notation, the tensors $\vec{\mathcal{A}}_{0}$,
$\vec{\Gamma}_{0}$, and $\vec{\mathcal{K}}_{0}$ are defined by
\begin{equation}
\mathcal{A}_{0jisk}=J^{-1} F_{j \alpha} F_{s \beta}
\mathcal{A}_{\alpha i \beta k},\quad \Gamma_{0jik}=F_{j\alpha}
F_{\beta k}^{-1} \Gamma_{\alpha i \beta},\quad \mathcal{K}_{0ij}=J
F_{\alpha i}^{-1} F_{\beta j}^{-1} \mathcal{K}_{\alpha
\beta}\label{module_inst}
\end{equation}
for an unconstrained material. For an incompressible material
$J=1$ in the above and \eqref{dotT0Hl0} is replaced by
\begin{equation}
\vec{\dot{T}}_0=\vec{\mathcal{A}}_0\vec{d}+\vec{\Gamma}_0\vec{\dot{B}}_{l0}+p\vec{d}-\dot{p}\vec{I},\quad
\vec{\dot{H}}_{l0}=\vec{\Gamma}_0\vec{d}+\vec{\mathcal{K}}_0\vec{\dot{B}}_{l0},\label{dotT0Hl0-inc}
\end{equation}
and the incremental incompressibility condition is
\begin{equation}
\div\vec{u}=0.\label{incomp-incr}
\end{equation}

Notice that $\vec{\mathcal{A}}_0$ and $\vec{\mathcal{K}}_0$
inherit the symmetries of $\vec{\mathcal{A}}$ and
$\vec{\mathcal{K}}$, respectively, so that
\begin{equation}
\mathcal{A}_{0jisk}=\mathcal{A}_{0skji}, \quad \mathcal{K}_{0ij}
=\mathcal{K}_{0ji}.
\end{equation}

Finally, using the incremental form of the rotational balance
condition $\vec{FT}=(\vec{FT})^t$, we find that
$\vec{\Gamma}_0$ has the symmetry
\begin{equation}
\Gamma_{0ijk}=\Gamma_{0jik},
\end{equation}
and we uncover the connections
\begin{equation}
\mathcal{A}_{0jisk}-\mathcal{A}_{0ijsk}=\tau_{js}\delta_{ik}-\tau_{is}\delta_{jk}
\end{equation}
between the components of the tensors $\vec{\mathcal{A}}_0$ and
$\vec{\tau}$ for an unconstrained material (see, for example,
Ogden \cite{ogde1971} for the specialization of these in the
purely elastic case), and
\begin{equation}
\mathcal{A}_{0jisk}-\mathcal{A}_{0ijsk}=(\tau_{js}+p\delta_{js})\delta_{ik}-(\tau_{is}+p\delta_{is})\delta_{jk}
\end{equation}
for incompressible materials (see Chadwick \cite{chad1997} for the
elastic specialization).

Following Prikazchikov \cite{Prik04}, we decompose the tensor
$\vec{\mathcal{A}}_{0}$ into the sum
\begin{equation}
\vec{\mathcal{A}}_{0}=\vec{\mathcal{A}}^{(0)}_{0}
+\vec{\mathcal{A}}^{(5)}_{0}+\vec{\mathcal{A}}^{(6)}_{0}.
\end{equation}
The first term $\vec{\mathcal{A}}^{(0)}_{0}$ does not involve any
derivatives with respect to $I_4$, $I_5$, and $I_6$. Clearly, this
term is very similar to the tensor of elastic moduli associated
with isotropic elasticity in the absence of magnetic fields. In
component form it is given by
\begin{eqnarray}
 J  \mathcal{A}^{(0)}_{0j i s k} =&&
  4 b_{i j} b_{k s} \Omega_{11}
  + 4 \mathcal{N}_{i j} \mathcal{N}_{k s} \Omega_{22}
  + 4 J^4
        \delta_{i j} \delta_{k s } \Omega_{33}
  + 4 (b_{k s} \mathcal{N}_{i j} +b_{i j} \mathcal{N}_{k
  s})\Omega_{12}\nonumber\\[0.1cm]
  &&+ 4 (b_{k s} \delta_{i j} + b_{i j} \delta_{k s}
        )\Omega_{13}+ 4 J^2(\mathcal{N}_{k s} \delta_{i j} +  \mathcal{N}_{i
        j} \delta_{k s}) \Omega_{23}\nonumber\\[0.1cm]
        &&+2 \delta_{ik} b_{j s}  \Omega_1+2(
      2  b_{i j} b_{k s} +  \delta_{ik}
        \mathcal{N}_{j s}-  b_{j k }b_{i s} -
        b_{i k } b_{j s}) \Omega_2\nonumber\\[0.1cm]
        &&+2 J^2(2  \delta_{i j}
        \delta_{k s} -\delta_{i s} \delta_{ j k}) \Omega_3,
\end{eqnarray} where
\begin{equation}
\mathcal{N}_{ij}=b_{kk} b_{ij}-b_{ik}b_{kj}
\end{equation}
and $b_{ij}$ are the components of $\vec{b}$.

The terms $\mathcal{A}^{(5)}_{0jisk}$ and
$\mathcal{A}^{(6)}_{0jisk}$ may be expressed in the forms
\begin{eqnarray}
\mathcal{A}^{(5)}_{0jisk}=&&\mathcal{A}^{0(5)}_{0jisk} \Omega_5
+\sum_{m=1,\,m\neq 4}^{6} \mathcal{A}^{m(5)}_{0jisk}
\Omega_{m5},\nonumber\\
\mathcal{A}^{(6)}_{0\alpha i \beta j}=&&\mathcal{A}^{0(6)}_{0jisk}
\Omega_6 +\sum_{m=1,\,m\neq 4}^{6} \mathcal{A}^{m(6)}_{0jisk}
\Omega_{m6},
\end{eqnarray}
where $\mathcal{A}^{4(5)}_{0jisk} = 0$ and
\begin{align}
&  \mathcal{A}^{0(5)}_{0jisk} =2 J^{-1} a_j a_s \delta_{ik},
&& \mathcal{A}^{1(5)}_{0jisk} =4  J^{-1} ( a_k a_s b_{i j} + a_i a_j b_{k s}),
\notag \\[0.1cm]
&  \mathcal{A}^{2(5)}_{0jisk} =4  J^{-1} (a_k a_s \mathcal{N}_{ij}
  + a_i a_j \mathcal{N}_{ks}),
&& \mathcal{A}^{3(5)}_{0jisk} =4 J ( a_k a_s \delta_{ij} + a_i a_j
\delta_{k s}),
\notag \\[0.1cm]
&  \mathcal{A}^{5(5)}_{0jisk} =4 J^{-1} a_j a_i a_s a_k, &&
\mathcal{A}^{6(5)}_{0jisk}
 =2 J^{-1} (a_j a_i \mathcal{H}_{ks} + a_{k} a_s \mathcal{H}_{i j}),
\end{align}
with
\begin{equation}
\mathcal{H}_{ij}=a_j a_k b_{ik} + a_i a_k b_{jk}, \quad a_i=F_{i
\alpha} B_{l_\alpha}.
\end{equation}
Similarly, $\mathcal{A}^{4(6)}_{0j i s k}=0$ and
\begin{eqnarray}
\mathcal{A}^{0(6)}_{0j i s k} =&& 2
 J^{-1} (\delta_{ik} \mathcal{H}_{j s} +
   a_i a_s b_{j k} + a_j a_k b_{is} + a_j
   a_s b_{ik}+a_i a_k b_{j s}),
\notag \\[0.1cm]
\mathcal{A}^{1(6)}_{0j i s k}=&&
 4 J^{-1} (b_{k s} \mathcal{H}_{ij} + b_{i j} \mathcal{H}_{ks}
 ),\quad
 \mathcal{A}^{2(6)}_{0j i s k}=4 J^{-1}
   (\mathcal{H}_{ij} \mathcal{N}_{k s} +\mathcal{H}_{ks} \mathcal{N}_{i j}),
\notag \\[0.1cm]
\mathcal{A}^{3(6)}_{0j i s k}=&& 4 J (\mathcal{H}_{ks} \delta_{ij}
+\mathcal{H}_{ij} \delta_{k s}),\quad  \mathcal{A}^{5(6)}_{0j i s
k}=2 J^{-1} (a_i a_j \mathcal{H}_{ks} +
   a_k a_s \mathcal{H}_{ij}), \notag\\[0.1cm]
 \mathcal{A}^{6(6)}_{0j i s k}
 =&&4 J^{-1} \mathcal{H}_{ij} \mathcal{H}_{ks}.
\end{eqnarray}

The tensor $\vec{\Gamma}_{0}$ is decomposed as
\begin{equation}
\vec{\Gamma}_{0}=\vec{\Gamma}^{(1)}_{0}+\vec{\Gamma}^{(2)}_{0}+
\vec{\Gamma}^{(3)}_{0}+
\vec{\Gamma}^{(5)}_{0}+\vec{\Gamma}^{(6)}_{0},
\end{equation}
with components given by
\begin{eqnarray}
  \Gamma^{(1)}_{0j i k}=&&4 b_{ij}\mathcal{M}_{1k}
  , \quad
 \Gamma^{(2)}_{0j i k} =4
 \mathcal{N}_{ij}\mathcal{M}_{2k}
 , \quad
 \Gamma^{(3)}_{0j i k} =4 J^{2} \delta_{ij}
 \mathcal{M}_{3k}, \notag \\[0.1cm]
 \Gamma^{(5)}_{0j i k} =&&4 a_j a_i\mathcal{M}_{5k}+ 2 (a_j
  \delta_{i k}+a_i \delta_{j k}) \Omega_5, \notag \\[0.1cm]
 \Gamma^{(6)}_{0j i k} =&& 4 \mathcal{H}_{i j}\mathcal{M}_{6k} +2
 ( \delta_{ik} a_s b_{j s}+ a_i b_{j k}
  +\delta_{j k}a_s b_{i s} +a_{j}   b_{i k} ) \Omega_6,
\end{eqnarray}
where
\begin{equation}
\mathcal{M}_{i k}=F_{\alpha k}^{-1} B_{l_\alpha}  \Omega_{i4} +
a_k\Omega_{i5} + a_j b_{jk} \Omega_{i6}.
\end{equation}

Finally, we represent $\vec{\mathcal{K}}_0$  in the form
\begin{equation}
\vec{\mathcal{K}}_{0}=\vec{\mathcal{K}}^{(4)}_{0}+
\vec{\mathcal{K}}^{(5)}_{0}+\vec{\mathcal{K}}^{(6)}_{0},
\end{equation}
with components
\begin{eqnarray}
\mathcal{K}^{(4)}_{0i j}=&&2 J F^{-1}_{\alpha i}(2 B_{l_\alpha}
\mathcal{M}_{4 j}+ F^{-1}_{\alpha j} ) \Omega_4, \notag \\[0.1cm]
\mathcal{K}^{(5)}_{0i j}=&& 2 J( 2a_{i} \mathcal{M}_{5 j}+ \delta_{i j} \Omega_5 ),\notag
\\[0.1cm]
\mathcal{K}^{(6)}_{0i j}=&& 2 J( 2a_k b_{i k} \mathcal{M}_{6 j}+
b_{i j} \Omega_6).
\end{eqnarray}

For an incompressible material, the above expressions are
unaltered except that $J=1$ and all the terms $\Omega_3$ and
$\Omega_{n3}$, $n=1, \ldots, 6$ in $\mathcal{A}_{0ijkl}$,
$\Gamma_{0ijk}$, $\mathcal{K}_{0ij}$ are omitted.

\subsection{Outside the material}

The standard relation $\vec{B}=\mu_0\vec{H}$ in vacuum is
incremented to
\begin{equation}
\vec{\dot{B}}^\star=\mu_0 \vec{\dot{H}}^\star,\label{dotBHstar}
\end{equation}
where $\vec{\dot{B}}^\star$ and $\vec{\dot{H}}^\star$ are the
increments of $\vec{B}^\star$ and $\vec{H}^\star$, respectively.
These fields satisfy Maxwell's equations
\begin{equation}
\div \vec{\dot{B}}^\star=0,\quad\curl
\vec{\dot{H}}^\star=\vec{0}.\label{dotdivBstar}
\end{equation}
Finally, we increment the Maxwell stress of \eqref{taustar} to
\begin{equation} \label{taustardot}
\vec{\dot{\tau}}^\star =\mu_0^{-1}[\vec{\dot{B}}^\star \otimes
\vec{B}^\star+\vec{B}^\star \otimes \vec{\dot{B}}^\star -
(\vec{B}^\star \cdot \vec{\dot{B}}^\star) \vec{I}],
\end{equation}
noting that $\div\vec{\dot{\tau}}^\star=\vec{0}$.

\subsection{Incremental boundary conditions} \label{par_inc_BC}

At the boundary of the material, in addition to any applied
traction $\vec{t}_{\text a}$ (defined per unit reference area),
there will in general be a contribution from the Maxwell stress
exterior to the material.  This is a traction
$\vec{\tau}^\star\vec{n}$ per unit current area and can be `pulled
back' to the reference configuration to give a traction
$J\vec{\tau}^\star\vec{F}^{-t}\vec{N}$ per unit
reference area, in which case the boundary condition \eqref{T-bc}
is modified to
\begin{equation}
\vec{T}^t \vec{N}=J\vec{\tau}^\star\vec{F}^{-t}\vec{N}+\vec{t}_{\text
a}.
\end{equation}
On taking the increment of this equation, we obtain
\begin{equation}
\vec{\dot{T}}^t \vec{N}=J\vec{\dot{\tau}}^\star\vec{F}^{-t}\vec{N}
-J\vec{\tau}^\star\vec{F}^{-t}\vec{\dot{F}}^t \vec{F}^{-t}
+\dot{J}\vec{\tau}^\star\vec{F}^{-t}\vec{N}+\vec{\dot{t}}_{\text
a},
\end{equation}
and hence, on updating this from the reference configuration to
the current configuration,
\begin{equation} \label{T0_BC}
\vec{\dot{T}}_0^t\vec{n}=
\vec{\dot{\tau}}^\star\vec{n}
-\vec{\tau}^\star\vec{d}^t \vec{n}+(\div\vec{u})\vec{\tau}^\star\vec{n}
+\vec{\dot{t}}_{\text a}.
\end{equation}

Proceeding in a similar fashion for the other fields, we increment
the magnetic boundary conditions \eqref{BH-bcs-lagrangian} to
give, again after updating,
\begin{equation} \label{B0_BC}
(\vec{\dot{B}}_{l0} + \vec{d} \vec{B}^\star
-(\div\vec{u})\vec{B}^\star- \vec{\dot{B}}^\star) \cdot \vec{n}=0
\end{equation}
and
\begin{equation} \label{H0_BC}
(\vec{\dot{H}}_{l0}-\vec{d}^t \vec{H}^\star -
\vec{\dot{H}}^\star) \times \vec{n}=\vec{0}.
\end{equation}

\section{Pure homogeneous deformation of a half-space\label{homog}}

Here we summarize the basic equations for the pure homogeneous
deformation of a half-space in the presence of a magnetic field
normal to its boundary prior to considering a superimposed
incremental deformation in Section 5.

\subsection{The deformed half-space}

Let $X_1$, $X_2$, $X_3$ be rectangular Cartesian coordinates in
the undeformed half-space $\mathcal{B}_0$ and take $X_2=0$ to be
the boundary $\partial\mathcal{B}_0$, with the material occupying
the domain $X_2\geq 0$. In order to minimize the number of
parameters, we consider the material to be incompressible and
subject to a plane strain in the $(X_1,X_2)$ plane. With respect
to the Cartesian axes, the deformation is then defined by
$x_1=\lambda X_1,x_2=\lambda^{-1}X_2,x_3=X_3$. The components of
the deformation gradient tensor $\vec{F}$ and the right
Cauchy-Green tensor $\vec{c}$ are written $\mathsf{F}$ and
$\mathsf{c}$, respectively, and are given by
\begin{equation}
\mathsf{F}=\begin{bmatrix}
\lambda &0 & 0 \\
0 & \lambda^{-1} & 0 \\
0 & 0 & 1
\end{bmatrix},\quad
\mathsf{c}=\begin{bmatrix}
\lambda^2 & 0 & 0 \\
0 &\lambda^{-2} & 0 \\
0 & 0 &1 \\
\end{bmatrix},
\label{F_BP}
\end{equation}
where $\lambda$ is the principal stretch in the $X_1$ direction.
The invariants $I_1$ and $I_2$ are therefore
\begin{equation} \label{inv12_BP}
I_1=I_2=1+\lambda^2+\lambda^{-2}.
\end{equation}

We take the magnetic induction vector $\vec{B}$ to be in the $x_2$
direction and to be independent of $x_1$ and $x_3$.  It then
follows from $\div\vec{B}=0$ that its component $B_2$ is constant.
Thus,
\begin{equation}
B_{1}=0, \quad B_{2}\neq 0, \quad B_{3}=0.
\end{equation}
The associated Lagrangian field $\vec{B}_l=\vec{F}^{-1}\vec{B}$
then has components
\begin{equation}
B_{l1}=0, \quad B_{l2}=\lambda B_{2}, \quad B_{l3}=0,\label{B_BP}
\end{equation}
and the invariants involving the magnetic field are
\begin{equation}
I_4=B_{l2}^2, \quad I_5=\lambda^{-2} I_4, \quad I_6=\lambda^{-4}
I_4.\label{inv46_BP}
\end{equation}

We may now compute the stress field using \eqref{tau_iso},
\eqref{F_BP} and \eqref{B_BP}. The resulting non-zero components
of $\vec{\tau}$ are
\begin{align}
&\tau_{11}=2 \Omega_1 \lambda^2 +  2 \Omega_2 (\lambda^2 + 1) - p, \notag
\\[0.1cm]
&\tau_{22}= 2 \Omega_1 \lambda^{-2} + 2 \Omega_2 (1 + \lambda^{-2})
- p + 2 \Omega_5 \lambda^{-2} I_4 + 4 \Omega_6 \lambda^{-4} I_4,
\notag\\[0.1cm]
&\tau_{33}=2 \Omega_1 + 2 \Omega_2 (\lambda^2 + \lambda^{-2}) - p. \label{compon_tau}
\end{align}
The magnetic field $\vec{H}$ has components given by \eqref{H_iso}
as
\begin{equation}
H_1=0, \quad H_2=2(\Omega_4+\lambda^{-2} \Omega_5 + \lambda^{-4}
\Omega_6) \lambda B_{l2}, \quad H_3=0.
\end{equation}
Since $B_{l2}$ and $\lambda$ are constant, all the fields are
uniform and the equilibrium equations and Maxwell's equations are
satisfied.

In view of \eqref{inv12_BP} and \eqref{inv46_BP}, there are only
two independent variables, $\lambda$ and $I_4$. We thus introduce
a specialization $\omega(\lambda,I_4)$ of the total energy
$\Omega$, by the definition
\begin{equation}
\omega(\lambda,I_4)= \Omega(1+\lambda^2+\lambda^{-2},
1+\lambda^2+\lambda^{-2}, I_4, \lambda^{-2} I_4, \lambda^{-4}
I_4),\label{omega}
\end{equation}
from which it follows that
\begin{align}
&\omega_\lambda=2 \lambda^{-1} [(\lambda^2-\lambda^{-2})(\Omega_1
+\Omega_2) - \lambda^{-2} I_4 \Omega_5 - 2 \lambda^{-4} I_4 \Omega_6],
\notag\\[0.1cm]
&\omega_4=\Omega_4 + \lambda^{-2} \Omega_5 + \lambda^{-4} \Omega_6, \label{derivee_omega}
\end{align}
where $\omega_\lambda=\partial \omega / \partial \lambda,
\omega_4=\partial \omega / \partial I_4$. Hence,
\begin{equation}
\tau_{11} - \tau_{22}=\lambda \omega_\lambda, \quad H_2=2 \lambda
B_{l2} \omega_4.
\end{equation}

\subsection{Outside the material}

From the boundary conditions \eqref{BH-bcs-eulerian} applied at
the interface $x_2=X_2=0$, we have $B_2^\star=B_2$ and
$H_1^\star=H_3^\star=0$, while from \eqref{BHstar} it follows that
$B_1^\star=B_3^\star=0$ and $H_2^\star=\mu_0^{-1}
B_2^\star=\mu_0^{-1} B_2$. Outside the material we take the
magnetic field to be uniform and equal to its interface value,
Maxwell's equations are then satisfied identically,
$\vec{B}^\star$ therefore has components
\begin{equation}
B_1^\star=0, \quad B_2^\star=B_2=\lambda^{-1} B_{l2}, \quad
B_3^\star=0,
\end{equation}
and $\vec{H}^\star$ has components
\begin{equation}
H_1^\star=0, \quad H_2^\star=\mu_0^{-1} B_2=\mu_0^{-1}
\lambda^{-1} B_{l2}, \quad H_3^\star=0.
\end{equation}

From these expressions, we deduce that the non-zero components of
the Maxwell stress \eqref{taustar} are given by
\begin{equation}
\tau_{11}^\star=-\tau_{22}^\star=-\tfrac{1}{2} \mu_0^{-1} B_2^2=
-\tfrac{1}{2} \mu_0^{-1} \lambda^{-2}
I_4=\tau_{33}^\star.\label{taustar_BV}
\end{equation}
The applied mechanical traction on $x_2=0$ required to maintain
the plane strain deformation has a single non-zero component
$\tau_{22}-\tau_{22}^\star$.

\section{Surface stability} \label{section_surf_stab}

We now address the question of surface stability for the deformed
half-space by establishing a bifurcation criterion based on the
incremental static solution of the boundary-value problem. Biot
\cite{Biot65} initiated this approach, which has since been
successfully applied to a great variety of boundary-value
problems; see Ogden \cite{OgFu01} for pointers to the vast
literature on the subject.

\subsection{Magnetoelastic moduli}

First we note that since $F_{ij}=0$ for $i \neq j$ and
$B_{l1}=B_{l3}=0$ several simplifications occur in the expressions
for the components of the magnetoelastic moduli tensors
$\vec{\mathcal{A}}_0,\,\vec{\Gamma}_0,\,\vec{\mathcal{K}}_0$. In
particular, we have
\begin{align}
& \mathcal{A}_{0iijk}=0, \qquad \mathcal{K}_{0ij}=0,\quad
\mbox{for} \quad j\neq k, \notag
\\[0.1cm]
& \Gamma_{0ii3}= \Gamma_{03ii}=\Gamma_{0ii1}=\Gamma_{01ii}=0,  \notag \\[0.1cm]
& \Gamma_{0ijk}=0, \quad \mbox{for} \quad i\neq j \neq k \neq i.
\label{modzero}
\end{align}
For subsequent use we compute the quantities
\begin{eqnarray}
a=&&\mathcal{A}_{01212}, \quad
  2b=\mathcal{A}_{01111}+\mathcal{A}_{02222} - 2\mathcal{A}_{01221}
   - 2\mathcal{A}_{01122}, \quad
  c=\mathcal{A}_{02121},
\notag \\[0.1cm]
d=&&\Gamma_{0211}, \quad
  e=\Gamma_{0222}-\Gamma_{0112},\quad f=\mathcal{K}_{011}, \quad
  g=\mathcal{K}_{022}.\label{def_abcdefg}
\end{eqnarray}
Explicitly, we obtain
\begin{align}
&a=2 \lambda^2 (\Omega_1+\Omega_2)+2 I_4 \Omega_6,\notag \\[0.1cm]
&b=(\lambda^2+\lambda^{-2})(\Omega_1+\Omega_2) +  I_4
[ \lambda^{-2} \Omega_5 + (6 \lambda^{-4} -2) \Omega_6]\notag \\[0.1cm]
 &\quad \quad  +2(\lambda^4+\lambda^{-4}-2)(\Omega_{11}+2\Omega_{12}
 + \Omega_{22})  \notag\\[0.1cm]
& \quad \quad +4 I_4(\lambda^{-4}-1)[\Omega_{15} + \Omega_{25}
+ 2 \lambda^{-2} ( \Omega_{16} + \Omega_{26})] \notag \\[0.1cm]
& \quad \quad +2 I_4^2 \lambda^{-4} (\Omega_{55} + 4 \lambda^{-2}
\Omega_{56} + 4 \lambda^{-4} \Omega_{66}),\notag \\[0.1cm]
&c=2 \lambda^{-2} (\Omega_1 + \Omega_2) + 2 I_4 [\lambda^{-2}
\Omega_5 + (2 \lambda^{-4} + 1) \Omega_6],\notag \\[0.1cm]
&d=2 B_{l2} \lambda [\lambda^{-2} \Omega_5 +(\lambda^{-4}+1)\Omega_6],\notag \\[0.1cm]
&e=4 B_{l2} \lambda^{-1} [\Omega_5 +2 \lambda^{-2} \Omega_6
+ (1-\lambda^4)(\Omega_{14}+\Omega_{24})\notag \\[0.1cm]
&\quad \quad  + (\lambda^{-2} - \lambda^2)(\Omega_{15} +
\Omega_{25})
+ (\lambda^{-4} -1)(\Omega_{16} + \Omega_{26}) \notag \\[0.1cm]
& \quad \quad   + I_4(\Omega_{54} + \lambda^{-2} \Omega_{55} + 2
\lambda^{-2} \Omega_{46} + 3 \lambda^{-4}
\Omega_{56} + 2 \lambda^{-6} \Omega_{66})],\notag \\[0.1cm]
&f=2(\lambda^{-2} \Omega_4 + \Omega_5 + \lambda^2 \Omega_6) \notag \\[0.1cm]
&g=2(\lambda^2 \Omega_4 + \Omega_5 + \lambda^{-2} \Omega_6)
+ 4 I_4 (\lambda^2 \Omega_{44} + 2 \Omega_{45} + 2 \lambda^{-2} \Omega_{46} \notag \\[0.1cm]
& \phantom{123456} + \lambda^{-2} \Omega_{55} + 2 \lambda^{-4}
\Omega_{56} + \lambda^{-6} \Omega_{66}). \label{abcdefg}
\end{align}
In terms of the energy density $\omega(\lambda, I_4)$ we have the
connections
\begin{equation}
a - c    = \lambda \omega_\lambda,\quad 2(b + c) = \lambda^2
\omega_{\lambda \lambda}, \quad  e = -2 B_{l2} \lambda^2
\omega_{\lambda4}, \quad  g =  2 \lambda^2 (\omega_4 + 2 I_4
\omega_{44}),
\end{equation}
where $\omega_{\lambda \lambda}=\partial^2 \omega/ \partial
\lambda^2$, $\omega_{\lambda4}=\partial^2 \omega / \partial
\lambda \partial I_4$ and $\omega_{44}= \partial^2 \omega/
\partial I_4^2$.

\subsection{Incremental fields and equations}

We seek incremental solutions depending only on the in-plane
variables $x_1$ and $x_2$ such that $u_3=0$ and $\dot{B}_{l03}=0$.
Hence $u_i=u_i (x_1,x_2)$ and
$\dot{B}_{l0i}=\dot{B}_{l0i}(x_1,x_2)$ for $i=1,2$ and
$\dot{p}=\dot{p}(x_1,x_2)$.  In the following, a subscripted comma
followed by an index $i$ signifies partial differentiation with
respect to $x_i,\,i=1,2$.

The incremental version \eqref{incomp-incr} of the
incompressibility constraint reduces here to
\begin{equation}
u_{1,1} + u_{2,2}=0,\label{uii_BP}
\end{equation}
and hence there exists a function $\psi=\psi(x_1,x_2)$ such that
\begin{equation}
u_1=\psi_{,2}, \quad u_2=-\psi_{,1}.\label{upsi}
\end{equation}

Similarly, equation \eqref{cond3delta0}$_3$ reduces to
\begin{equation}
\dot{B}_{l01,1}+\dot{B}_{l02,2}=0,\label{Bii_BP}
\end{equation}
and the function $\phi=\phi(x_1,x_2)$ is introduced such that
\begin{equation}
\dot{B}_{l01}=\phi_{,2}, \quad
\dot{B}_{l02}=-\phi_{,1}.\label{Bphi}
\end{equation}

The incremental equations of equilibrium \eqref{cond3delta0}$_1$
simplify to
\begin{equation}
\dot{T}_{011,1} + \dot{T}_{021,2}=0, \quad  \dot{T}_{012,1} +
\dot{T}_{022,2}=0.\label{dotT0rel_BP}
\end{equation}

From the identities \eqref{modzero}, the only non-zero components
of the incremental stress $\vec{\dot{T}}_0$ are found to be
\begin{align}
&\dot{T}_{011} = (
\mathcal{A}_{01111}+p) u_{1,1} + \mathcal{A}_{01122}
u_{2,2} +\dot{B}_{0_{2}} \Gamma_{0112}- \dot{p},\notag \\[0.1cm]
& \dot{T}_{021} =
(\mathcal{A}_{02112}+p) u_{2,1} + \mathcal{A}_{02121}
u_{1,2} +\dot{B}_{0_{1}} \Gamma_{0211},\notag \\[0.1cm]
&\dot{T}_{012} =
 (\mathcal{A}_{01221} +p) u_{1,2} +\mathcal{A}_{01212} u_{2,1}
 + \dot{B}_{0_{1}} \Gamma_{0121}, \notag \\[0.1cm]
&\dot{T}_{022} =(\mathcal{A}_{02222}+p) u_{2,2} +
\mathcal{A}_{02211} u_{1,1} +
 \dot{B}_{0_{2}} \Gamma_{0222}- \dot{p}.\label{dotT12_BP}
\end{align}

Also, equation \eqref{cond3delta0}$_2$ reduces to
\begin{equation}
\dot{H}_{l01,2} - \dot{H}_{l02,1}=0,\label{dotH12_BP}
\end{equation}
wherein are the only non-zero components of $\vec{\dot{H}}_l$,
which, from \eqref{modzero}, are given by
\begin{equation}
\dot{H}_{l01}=\Gamma_{0121} (u_{1,2} + u_{2,1}) +
\mathcal{K}_{011} \dot{B}_{l01}, \quad \dot{H}_{l02}=\Gamma_{0112}
u_{1,1} + \Gamma_{0222} u_{2,2} + \mathcal{K}_{022}
\dot{B}_{l02}.\label{dotH_BP}
\end{equation}

In terms of the functions $\psi$ and $\phi$ equations
\eqref{dotT0rel_BP} and \eqref{dotH12_BP} become
\begin{align}
&(\mathcal{A}_{01111} - \mathcal{A}_{01122} - \mathcal{A}_{01221})
\psi_{,112} + \mathcal{A}_{02121} \psi_{,222} - \Gamma_{0112} \phi_{,11}
+ \Gamma_{0121} \phi_{,22} = \dot{p}_{,1}, \notag \\[0.1cm]
&(\mathcal{A}_{02222} - \mathcal{A}_{01122} - \mathcal{A}_{01221})
\psi_{,122} + \mathcal{A}_{01212} \psi_{,111} - (\Gamma_{0121}
-\Gamma_{0222}) \phi_{,12} =- \dot{p}_{,2}, \notag  \\[0.1cm]
&(\Gamma_{0222} - \Gamma_{0112} - \Gamma_{0121}) \psi_{,112} +
\Gamma_{0121} \psi_{,222} + \mathcal{K}_{022} \phi_{,11} +
\mathcal{K}_{011} \phi_{,22}=0.\label{equil-incre}
\end{align}
We eliminate $\dot{p}$ from the first two equations by
cross-differentiation and addition and obtain finally the coupled
equations
\begin{equation} \label{eqmov1}
a \psi_{,1111} + 2b \psi_{,1122} + c \psi_{,2222} + (e-d)
\phi_{,112} + d \phi_{,222}=0
\end{equation}
and
\begin{equation} \label{eqmov2}
d \psi_{,222} + (e-d) \psi_{,112} + f \phi_{,22} + g \phi_{,11}=0
\end{equation}
for $\psi$ and $\phi$.

\subsection{Outside the material}

In vacuum, Maxwell's equations \eqref{dotdivBstar} hold for
$\vec{\dot{B}}$ and $\vec{\dot{H}}$. From the second equation, and
the assumption that all fields depend only on $x_1$ and $x_2$, we
deduce the existence of a scalar function
$\phi^\star=\phi^\star(x_1,x_2)$ such that
\begin{equation}
\dot{H}_1^\star=-\phi_{,1}^\star, \quad
\dot{H}_2^\star=-\phi_{,2}^\star, \quad
\dot{H}_3^\star=0.\label{Hstarphi}
\end{equation}
Equation \eqref{dotBHstar} then gives
\begin{equation}
\dot{B}_1^\star=-\mu_0 \phi_{,1}^\star, \quad \dot{B}_2^\star=-\mu_0 \phi_{,2}^\star, \quad \dot{B}_3^\star=0,
\end{equation}
and from \eqref{dotdivBstar}$_1$ we obtain the equation
\begin{equation}
\phi^\star_{,11}+\phi^\star_{,22}=0\label{phi1122}
\end{equation}
for $\phi^\star$. Finally, the incremental Maxwell stress tensor
\eqref{taustardot} has non-zero components
\begin{equation}
\dot{\tau}_{11}^\star=\lambda^{-1} B_{l2}
\phi_{,2}^\star=\dot{\tau}_{33}^\star=-\dot{\tau}_{22}^\star,
\quad \dot{\tau}_{12}^\star=-\lambda^{-1} B_{l2}
\phi_{,1}^\star=\dot{\tau}_{21}^\star.\label{incrmax}
\end{equation}

\subsection{Boundary conditions}

We now specialize the general incremental boundary conditions of
Section \ref{par_inc_BC} to the present deformed semi-infinite
solid. First, for $\vec{\dot{t}}_{\text a}=\vec{0}$, the
incremental traction boundary conditions \eqref{T0_BC} reduce to
\begin{equation}
\dot{T}_{021} + \tau^\star_{11} u_{2,1}-\dot{\tau}_{21}^\star=0,
\quad \dot{T}_{022} + \tau_{22}^\star u_{2,2} -
\dot{\tau}_{22}^\star=0,\label{incremental-traction-halfspace}
\end{equation}
on $x_2=0$. Putting together the results of this section, using
\eqref{taustar_BV}, \eqref{def_abcdefg}, \eqref{upsi},
\eqref{Bphi}, \eqref{dotT12_BP}, \eqref{Hstarphi} and
\eqref{incrmax}, we express the two equations
\eqref{incremental-traction-halfspace} as
\begin{equation} \label{eq1_BC}
(\tau_{22} + \textstyle{\frac{1}{2}} \mu_0^{-1} \lambda^{-2} I_4
-c) \psi_{,11} + c \psi_{,22} + d \phi_{,2} + \lambda^{-1} B_{l2}
\phi^\star_{,1}=0,
\end{equation}
and
\begin{equation} \label{eq2_BC}
(2b+c-\tau_{22} + \tfrac{1}{2} \mu_0^{-1} \lambda^{-2} I_4)
\psi_{,112} + c \psi_{,222} + e \phi_{,11} +d \phi_{,22} -
\lambda^{-1} B_{l2} \phi^\star_{,12}=0,
\end{equation}
which apply on $x_2=0$.  In obtaining the latter we have
differentiated \eqref{incremental-traction-halfspace}$_2$ with
respect to $x_1$ and made use of \eqref{equil-incre}$_1$.

Next, the incremental magnetic boundary conditions \eqref{B0_BC}
and \eqref{H0_BC} reduce to
\begin{equation}
\dot{B}_{l02} + B_2^\star u_{2,2} - \dot{B}^\star_2=0, \quad
\dot{H}_{l01} - H_2^\star u_{2,1} - \dot{H}_1^\star=0
\end{equation}
on $x_2=0$. Using again the results of the preceding sections, we
write these as
\begin{equation}
\lambda^{-1} B_{l2} \psi_{,12} + \phi_{,1} - \mu_0
\phi_{,2}^\star=0,\label{eq3_BC}
\end{equation}
and
\begin{equation}
(\mu_0^{-1} \lambda^{-1} B_{l2} -d) \psi_{,11} + d \psi_{,22} + f
\phi_{,2} + \phi^\star_{,1}=0\label{eq4_BC}
\end{equation}
on $x_2=0$.

\subsection{Resolution}

We are now in a position to solve the incremental boundary value
problem. We seek small-amplitude solutions, localized near the
interface $x_2=0$. Hence we take solutions in the solid ($x_2\geq
0$) to be of the form
\begin{equation} \label{psiphi}
\psi=A e^{-ksx_2} e^{\imi k x_1}, \quad  \phi=kDe^{-ksx_2} e^{\imi k x_1},
\end{equation}
where $k>0$ ($2 \pi/k$ is the wavelength of the perturbation) and
$s$ is such that
\begin{equation} \label{res}
\Re (s) >0
\end{equation}
to ensure decay with increasing $x_2\,(>0)$.

Substituting \eqref{psiphi} into the incremental equilibrium
equations \eqref{eqmov1} and \eqref{eqmov2}, we obtain
\begin{align}
&(cs^4 -2bs^2+a)A - s (ds^2 + d-e)D=0,\notag \\[0.1cm]
&s(ds^2 + d-e)A - (fs^2-g)D=0.\label{systAB}
\end{align}
For non-trivial solutions to exist, the determinant of
coefficients of $A$ and $D$ must vanish, which yields a cubic in
$s^2$, namely
\begin{equation}
(cf-d^2) s^6 - [2bf + cg + 2 (d-e)d] s^4 + [2bg + af - (d-e)^2]s^2
- ag=0.\label{cubic}
\end{equation}
From the six possible roots we select $s_1$, $s_2$, $s_3$ to be
the three roots satisfying \eqref{res}. We then construct the
general solution for the solid as
\begin{equation}
\psi=\sum_{j=1}^3 A_j e^{-ks_j x_2} e^{\imi k x_1}, \quad
\phi=k \sum_{j=1}^3 D_j e^{-ks_j x_2}e^{\imi k x_1},
\end{equation}
where $A_j,\,D_j,\,j=1,2,3$, are constants.

For the half-space $x_2\leq 0$ (vacuum) we take a solution
$\phi^\star$ to \eqref{phi1122} that is localized near the
interface $x_2=0$. Specifically, we write this as
\begin{equation} \label{phistar}
\phi^\star=\imi k C^\star e^{k x_2} e^{\imi k x_1},
\end{equation}
where $C^\star$ is a constant.

The constants $A_j$ and $D_j$ are related through either equation
in \eqref{systAB}. From the second equation, for instance, we
obtain
\begin{equation} \label{ABrel}
s_j (d s_j^2 + d-e) A_j + (f s_j^2 - g) D_j =0,\quad j=1,2,3;\
\mbox{no summation}.
\end{equation}
We also have the two traction boundary conditions \eqref{eq1_BC}
and \eqref{eq2_BC}, which read
\begin{multline}
(c-\tau_{22} - \tfrac{1}{2} \mu_0^{-1} \lambda^{-2} I_4)(A_1
+ A_2 + A_3) + c (s_1^2 A_1 + s_2^2 A_2 + s_3^2 A_3)\\[0.1cm]
 - d (s_1 D_1 + s_2 D_2 + s_3 D_3 ) - \lambda^{-1} B_{l2} C^\star=0,\label{relBC1}
\end{multline}
and
\begin{multline}
(\tau_{22} - \tfrac{1}{2} \mu_{0}^{-1} \lambda^{-2}
I_4 - 2b -c) (s_1 A_1 + s_2 A_2 + s_3 A_3) \\[0.1cm]
+ c(s_1^3 A_1 + s_2^3 A_2 + s_3^3 A_3) + (e-d s_1^2) D_1\\[0.1cm]
 + ( e-ds_2^2) D_2 + (e - d s_3^2) D_3 -  \lambda^{-1} B_{l2} C^\star=0. \label{relBC2}
\end{multline}
Finally, the two magnetic boundary conditions \eqref{eq3_BC} and
\eqref{eq4_BC} become
\begin{equation}
\lambda^{-1} B_{l2} (s_1 A_1 + s_2 A_2 + s_3 A_3) - (D_1 + D_2 +
D_3 ) + \mu_0 C^\star=0, \label{relBC3}
\end{equation}
and
\begin{multline}
(d-\mu_0^{-1} \lambda^{-1} B_{l2})(A_1+ A_2+ A_3) + d (s_1^2 A_1 + s_2^2 A_2 + s_3^2 A_3) \\[0.1cm]
- f (s_1 D_1 + s_2 D_2 + s_3 D_3) - C^\star=0.\label{relBC4}
\end{multline}

In total, there are seven homogeneous linear equations for the
seven unknowns $A_j$, $D_j,\,j=1,2,3$, and $C^\star$. The
resulting determinant of coefficients must vanish and this
equation is rather formidable to solve, particularly since it must
be solved in conjunction with the bicubic \eqref{cubic}. It is in
principle possible to express the determinant in terms of the sums
and products $s_1 + s_2 + s_3$, $s_1 s_2 + s_2 s_3 + s_3 s_1$,
$s_1 s_2 s_3$, and to find these from the bicubic \eqref{cubic},
similarly to the analysis conducted in the purely elastic case
(see Destrade \textit{et al.} \cite{DeOt05}). However, the
resulting algebraic expressions rapidly become too cumbersome for
this approach to be pursued.

Instead, we propose either
\begin{itemize}
\item[(a)] to turn directly to a numerical treatment once $\Omega$
has been determined by curve fitting from experimental data for a
given magnetoelastic solid,

or \item[(b)] to use a simple form for $\Omega$ that allows some
progress to be made.
\end{itemize}

Regarding approach (a), we remark that, as emphasized by Dorfmann
and Ogden \cite{DoOg04, DoOg03a, DoOg03b, OgDo05}, there is a
shortage of, and a pressing need for, suitable experimental data
and for the derivation of functions $\Omega$ from such data. In
the next section we focus primarily on the analytical approach
(b).

\subsection{Example: a ``Mooney-Rivlin magnetoelastic solid''}

As a prototype for the energy function $\Omega$, we propose
\begin{equation}
\Omega= \tfrac{1}{4}\mu(0) [(1+\gamma)(I_1-3) + (1-\gamma)(I_2-3)]
+ \mu^{-1}_0 (\alpha I_4 + \beta I_5),\label{MR}
\end{equation}
where $\mu(0)$ is the shear modulus of the material in the absence
of magnetic fields and $\alpha$, $\beta$, $\gamma$ are
dimensionless material constants, $\alpha$ and $\beta$ being
magnetoelastic coupling parameters. For $\alpha = \beta = 0$,
\eqref{MR} reduces to the strain energy of the elastic
Mooney-Rivlin material, a model often used for elastomers.

In respect of \eqref{MR} the stress $\vec{\tau}$ in
\eqref{tau_iso} reduces to
\begin{equation}
\vec{\tau}=- p \vec{I}+\tfrac{1}{2}\mu_0(1+\gamma) \vec{b} +
\tfrac{1}{2}\mu_0 (1-\gamma) ( I_1 \vec{b} - \vec{b}^2)  +
2\mu_0^{-1}\beta \vec{B} \otimes \vec{B} ,\label{tau-MR}
\end{equation}
while $\vec{H}$ in \eqref{H_iso} becomes
\begin{equation}
\vec{H}=2\mu_0^{-1}(\alpha \vec{b}^{-1} \vec{B} + \beta
\vec{B}).\label{H-MR}
\end{equation}
Clearly, equation \eqref{tau-MR} shows that the parameter $\alpha$
does not affect the stress. By contrast $\beta$, if positive,
stiffens the material in the direction of the magnetic field, i.e.
a larger normal stress in this direction is required to achieve a
given extension in this direction than would be the case without
the magnetic field. On the other hand, by reference to
\eqref{H-MR}, we see that $\alpha$ provides a measure of how the
magnetic properties of the material are influenced by the
deformation (through $\vec{b}$). If $\beta=0$ the stress is
unaffected by the magnetic field.  On the other hand, if
$\alpha=0$ then the magnetic constitutive equation \eqref{H-MR} is
unaffected by the deformation. Thus, a two-way coupling requires
inclusion of both constants.

The quantities defined in \eqref{def_abcdefg} and \eqref{abcdefg}
now reduce to
\begin{align}
&a  = \mu(0) \lambda^2, \qquad
 2b = \mu(0)(\lambda^2+\lambda^{-2} + \beta \lambda^{-2} \bar{I}_4), \qquad
 c  = \mu(0)(\lambda^{-2} + \beta \lambda^{-2} \bar{I}_4),
\notag \\[0.1cm]
&d = \sqrt{\mu_0^{-1} \mu(0)}\beta \lambda^{-1} \bar{B}_{l2},
\qquad
 e = 2 \sqrt{\mu_0^{-1} \mu(0)} \beta \lambda^{-1} \bar{B}_{l2}, \notag
 \\[0.1cm]
&f = \mu_0^{-1}(\alpha \lambda^{-2} + \beta),\qquad
 g = \mu_0^{-1}(\alpha \lambda^2+\beta),
\end{align}
where $\bar{B}_{l2}$, a dimensionless measure of the magnetic
induction vector amplitude, and $\bar{I}_4$ are defined by
\begin{equation}
\bar{B}_{l2}=B_{l2} / \sqrt{\mu_0 \mu(0)},\quad
\bar{I}_4=\bar{B}_{l2}^2.
\end{equation}
Note the connections
\begin{equation}
2b=a+c, \quad e=2d.
\end{equation}

Now we find that the bicubic \eqref{cubic} factorizes in the form
\begin{equation}
(s^2-1)(s^2-\lambda^4)[\alpha \lambda^4 + \beta \lambda^2 -(\alpha
+ \beta \lambda^2 + \alpha \beta \bar{I}_4)s^2]=0,
\end{equation}
and it follows that the relevant roots are
\begin{equation} \label{rootS}
s_1=1, \quad s_2=\lambda^2, \quad s_3 = \lambda \sqrt{\frac{\alpha
\lambda^2 + \beta}{\alpha + \beta  \lambda^2 + \alpha \beta
\bar{I}_4}}.
\end{equation}
Note that for $s_3$ to be real for all $\lambda>0$ and all
$\bar{B}_{l2}$, the inequalities
\begin{equation}
\alpha\geq 0, \quad \beta> 0\quad\text{or}\quad \alpha> 0, \quad
\beta\geq 0
\end{equation}
must hold.  (The case in which there is no magnetic field
corresponds to $\alpha=\beta=0$.)  It is assumed here that these
inequalities are satisfied, so that $s_3$ is indeed a qualifying
root satisfying \eqref{res}.

The  equation \eqref{ABrel} becomes
\begin{equation}
s_j (s_j^2-1)\beta \lambda^{-1} \bar{B}_{l2} \hat{A}_j  - [(\alpha
\lambda^{-2} + \beta)s_j^2 - \alpha \lambda^2-\beta]
\hat{D}_j=0,\quad j=1,2,3,\label{3first}
\end{equation}
where
\begin{equation}
\hat{A}_j=\sqrt{\mu^{-1}_0\mu(0)} A_j, \quad \hat{D}_j=\mu^{-1}_0
D_j,
\end{equation}
and the $s_j$ are given by \eqref{rootS}.

Next, consider the four remaining boundary conditions
\eqref{relBC1}--\eqref{relBC4}. In order to keep the number of
parameters to a minimum (so far, we have $\lambda$,
$\bar{B}_{l2}$, $\alpha$, $\beta$), and to make a simple
connection with known results for the surface stability of an
elastic Mooney-Rivlin material, we assume that there is no applied
mechanical traction on the boundary $x_2=0$, and hence
\begin{equation}
\tau_{22}=\tau_{22}^\star=\tfrac{1}{2} \mu_0^{-1} \lambda^{-2}
I_4.
\end{equation}

The boundary conditions \eqref{eq1_BC}--\eqref{eq4_BC} now read
\begin{align}
&[1+(\beta-1) \bar{I}_4](\hat{A}_1+\hat{A}_2+\hat{A}_3) +(1+\beta
\bar{I}_4)(s_1^2 \hat{A}_1+s_2^2 \hat{A}_2+s_3^2
\hat{A}_3)  \notag \\[0.1cm]
 & \qquad  \qquad \qquad \qquad \qquad \qquad -\beta \lambda \bar{B}_{l2}
 (s_1 \hat{D}_1 + s_2 \hat{D}_2 + s_3 \hat{D}_3) - \lambda \bar{B}_{l2} C^\star =0,  \notag \\[0.1cm]
&(\lambda^4 + 2 + 2 \beta \bar{I}_4) (s_1 \hat{A}_1 + s_2
\hat{A}_2 + s_3 \hat{A}_3) - (1+\beta \bar{I}_4) (s_1^3
\hat{A}_1 + s_2^3 \hat{A}_2 + s_3^3 \hat{A}_3) \notag \\[0.1cm]
& \qquad \qquad \qquad + \beta \lambda \bar{B}_{l2} [(s_1^2 -2)
\hat{D_1} + (s_2^2-2) \hat{D}_2 + (s_3^2-2) \hat{D}_3] + \lambda
\bar{B}_{l2} C^\star=0,  \notag \\[0.1cm]
&\bar{B}_{l2} (s_1 \hat{A}_1 + s_2 \hat{A}_2 + s_3 \hat{A}_3) -
\lambda (\hat{D}_1 + \hat{D}_2 + \hat{D}_3) + \lambda C^\star=0, \notag \\[0.1cm]
&\lambda \bar{B}_{l2} (\beta-1) (\hat{A}_1 + \hat{A}_2 +
\hat{A}_3) + \lambda \beta \bar{B}_{l2} (s_1^2 \hat{A}_1 + s_2^2
\hat{A}_2
+ s_3^2 \hat{A}_3)\notag \\[0.1cm]
&\qquad \qquad \qquad \qquad - (\alpha + \beta \lambda^2) (s_1
\hat{D}_1 + s_2 \hat{D}_2 + s_3 \hat{D}_3) -\lambda^2
C^\star=0.\label{4remaining}
\end{align}

From the seven equations \eqref{3first} and \eqref{4remaining}, we
have derived a bifurcation criterion (vanishing of the determinant
of coefficients) using a computer algebra package, but it is too
long to reproduce here. It is a complicated rational function of
the four parameters $\lambda$, $\bar{B}_{l2}$, $\alpha$, $\beta$.
However, it is easy to solve numerically, and for the numerical
examples we fix the material parameters $\alpha$ and $\beta$ and
find the critical stretch $\lambda_\text{cr}$ in compression as a
function of $\bar{B}_{l2}$. For $\bar{B}_{l2} = 0$, we recovered
the well-known critical compression stretch for surface
instability of the elastic Mooney-Rivlin material in plane strain,
namely $\lambda_\text{cr} = 0.5437$ \cite{Biot65}, as expected.
For Figure 1a (Figure 1b), we set $\alpha = 0.5$ ($\alpha = 2.0$)
and curves for $\beta = 0.0, 0.5, 1.0, 1.5, 2.0$ are shown. We
found that $\lambda_\text{cr}$ is an even function of
$\bar{B}_{l2}$ and we therefore restricted attention to positive
$\bar{B}_{l2}$ (within the range $0 \leq \bar{B}_{l2} \leq 3$).
The behaviour as $\bar{B}_{l2}$ becomes larger and larger
(not shown here) indicates that the half-space becomes more and
more unstable in compression. Moreover, it can even become
unstable in tension ($\lambda_\text{cr}
> 1$). The figures also clearly demonstrate that for some values
of $\alpha$, $\beta$, and $\bar{B}_{l2}$ the critical stretch
ratio is smaller than that for the purely elastic case
($\lambda_\text{cr} < 0.5437$), in which cases the magnetic field
has a stabilizing effect. 
\begin{figure}
\centering \subfigure[$\alpha = 0.5$]{\epsfig{figure=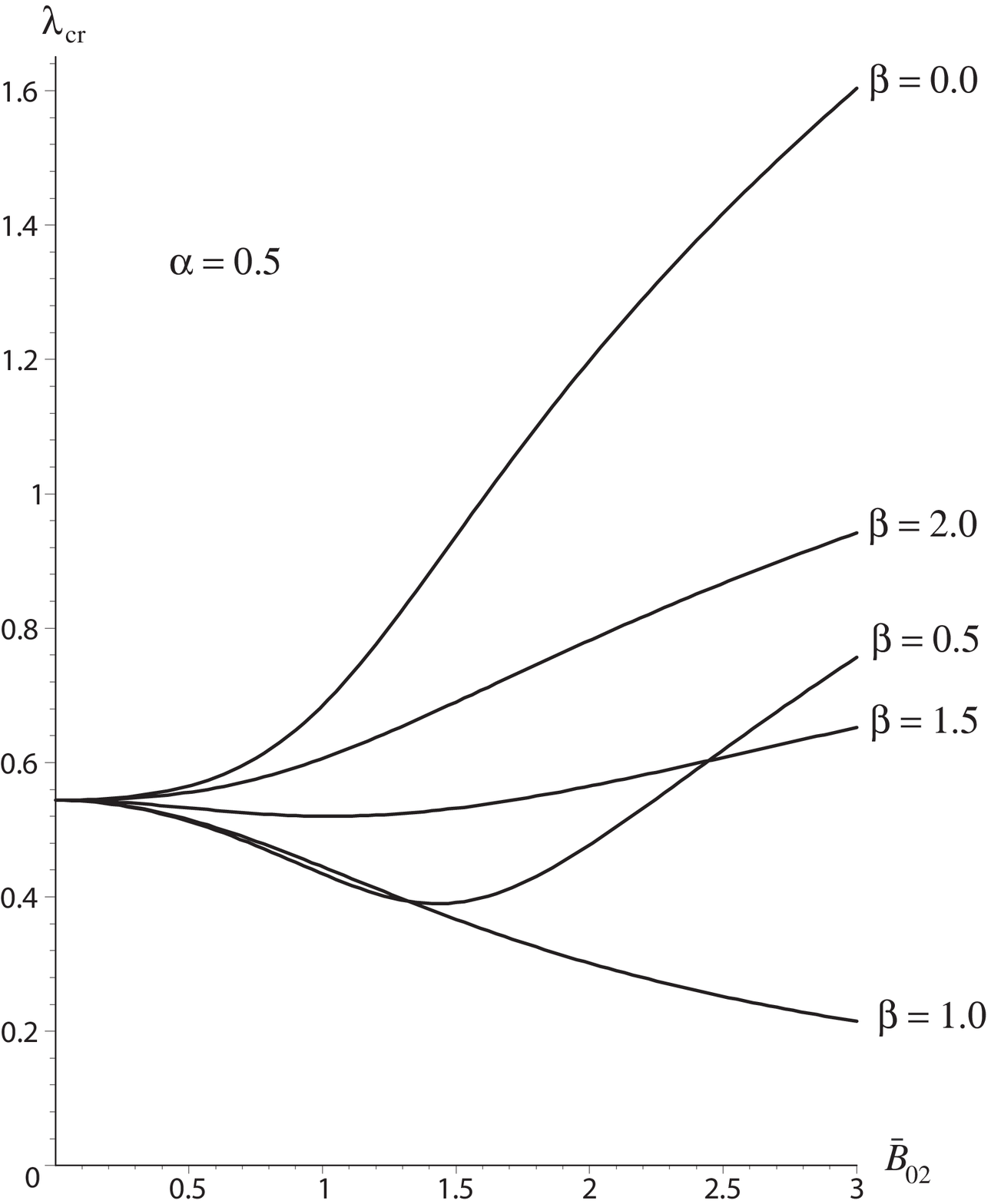,
width=.45\textwidth, height=.5\textwidth}}
  \quad
     \subfigure[$\alpha = 2.0$]{\epsfig{figure=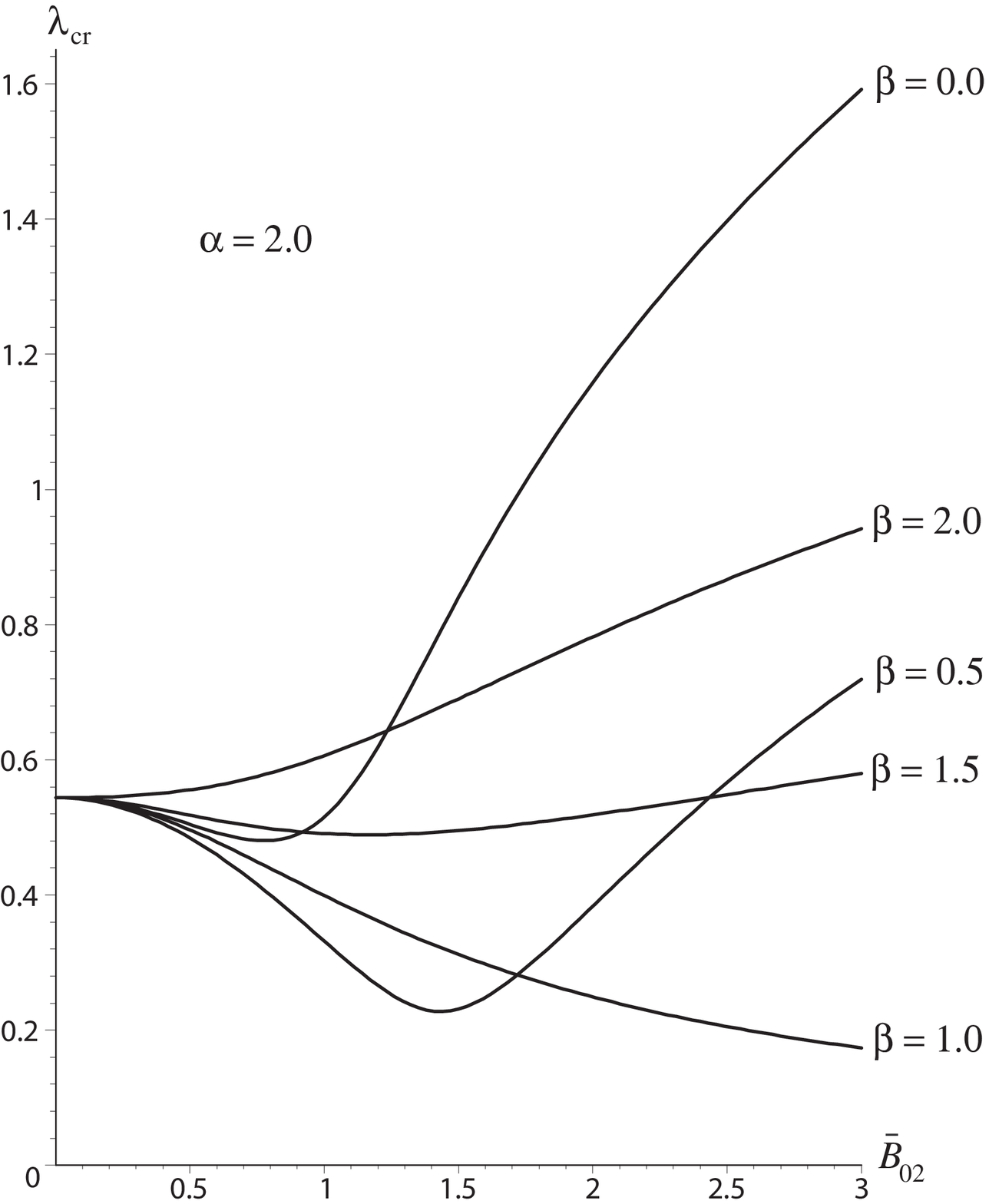,
width=.45\textwidth, height=.5\textwidth}} \caption{Dependence of
the critical stretch $\lambda_\text{cr} < 1$ for instability in
compression for a magnetoelastic Mooney-Rivlin solid in plane
strain on the non-dimensional measure $\bar{B}_{l2}$ of the
magnetic field for several values of the magnetoelastic
coupling parameters $\alpha$ and $\beta$.}
\end{figure}

Turning back to a phenomenological approach, we remark that the
energy function \eqref{MR} has quite good curve-fitting qualities
for moderate fields. There are four parameters at hand, namely
$\mu(0)$, $\alpha$, $\beta$, $\gamma$, two of which, $\mu(0)$ and
$\beta$, may be determined from shear tests. Indeed Dorfmann and
Ogden \cite{OgDo05} show that in general the shear modulus for
isotropic nonlinear magnetoelasticity is $2[\Omega_1 + \Omega_2 +
I_4 \Omega_5 + I_4 \Omega_6(3 + 2\kappa^2)]$, where $\kappa$ is
the amount of shear in a simple shear test. Here the modulus is
independent of $\kappa$ and is given by
\begin{equation}
\mu(B_{l2}) = \mu(0) + 2 \mu_0^{-1} \beta
I_4.\label{parabolic}
\end{equation}
This highlights the role of $\beta$ in increasing the
mechanical stiffness of the material --- through the shear
modulus. Jolly \textit{et al.} \cite{JoCa96} conducted double lap
shear tests on magneto-sensitive elastomers containing 10, 20, and
30\% by volume of iron particles. From their Figure 7, we see that
in the range $0 \leq B_{l2} \leq 0.5$ Tesla, the variations of
$\mu(B_{l2})$ resemble those of a parabolic profile such as the
one suggested by \eqref{parabolic}. For the 10\% iron by volume
elastomer specimen, Table 1 in Jolly \textit{et al.} \cite{JoCa96}
gives $\mu(0) = 0.26$ MPa, and at $B_{l2} = 0.5$ Tesla, we read
off their Figure 7 that $\mu(0.5) - \mu(0) \simeq 0.07$ MPa,
indicating that $\beta \simeq 0.18$. Similarly, for the
20\% and the 30\% iron by volume elastomer specimens we find
$\beta \simeq 0.53$ and $\beta \simeq 0.72$,
respectively.

Figure 2a (Figure 2b) illustrates the variation of the critical
compression stretch with the amplitude of the dimensional magnetic
induction vector, from 0 to 0.5 Tesla, for the 20\% (30\%) iron by
volume elastomer, and for several values of $\alpha$. We remark
than the presence of the magnetic field makes the two specimens
slightly more stable than in the purely elastic case because all
the critical compression stretch values are smaller than 0.5437.
It is also clear that increasing the value of $\alpha$ makes
the half-space more stable. However, it is worth noting that the
30\% iron by volume specimen is slightly less stable than the 20\%
iron by volume specimen for the same values of $\alpha$.

\begin{figure}
\centering \subfigure[$\beta = 0.53$]{\epsfig{figure=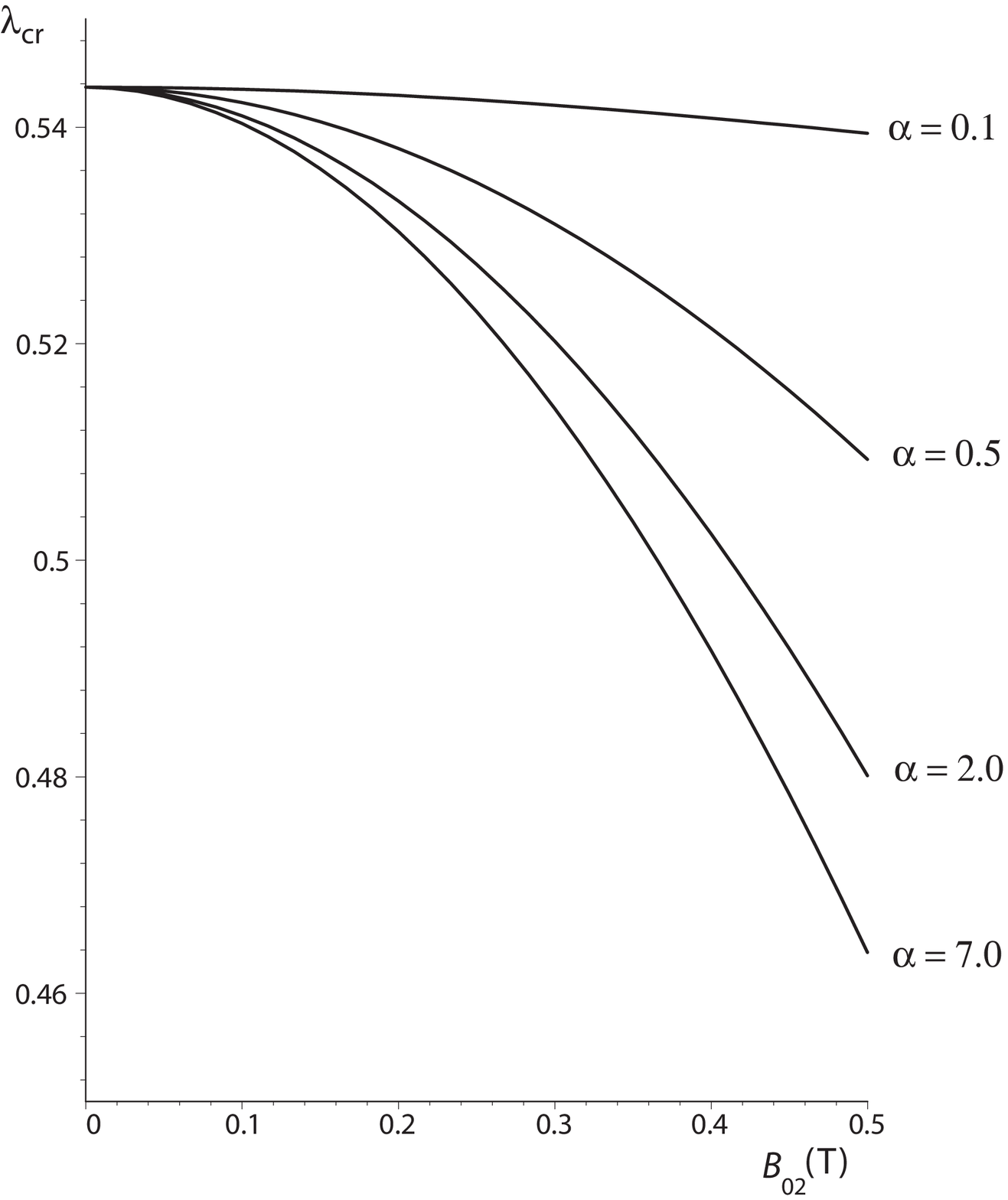,
width=.45\textwidth, height=.5\textwidth}}
 \quad
   \subfigure[ $\beta = 0.72$]{\epsfig{figure=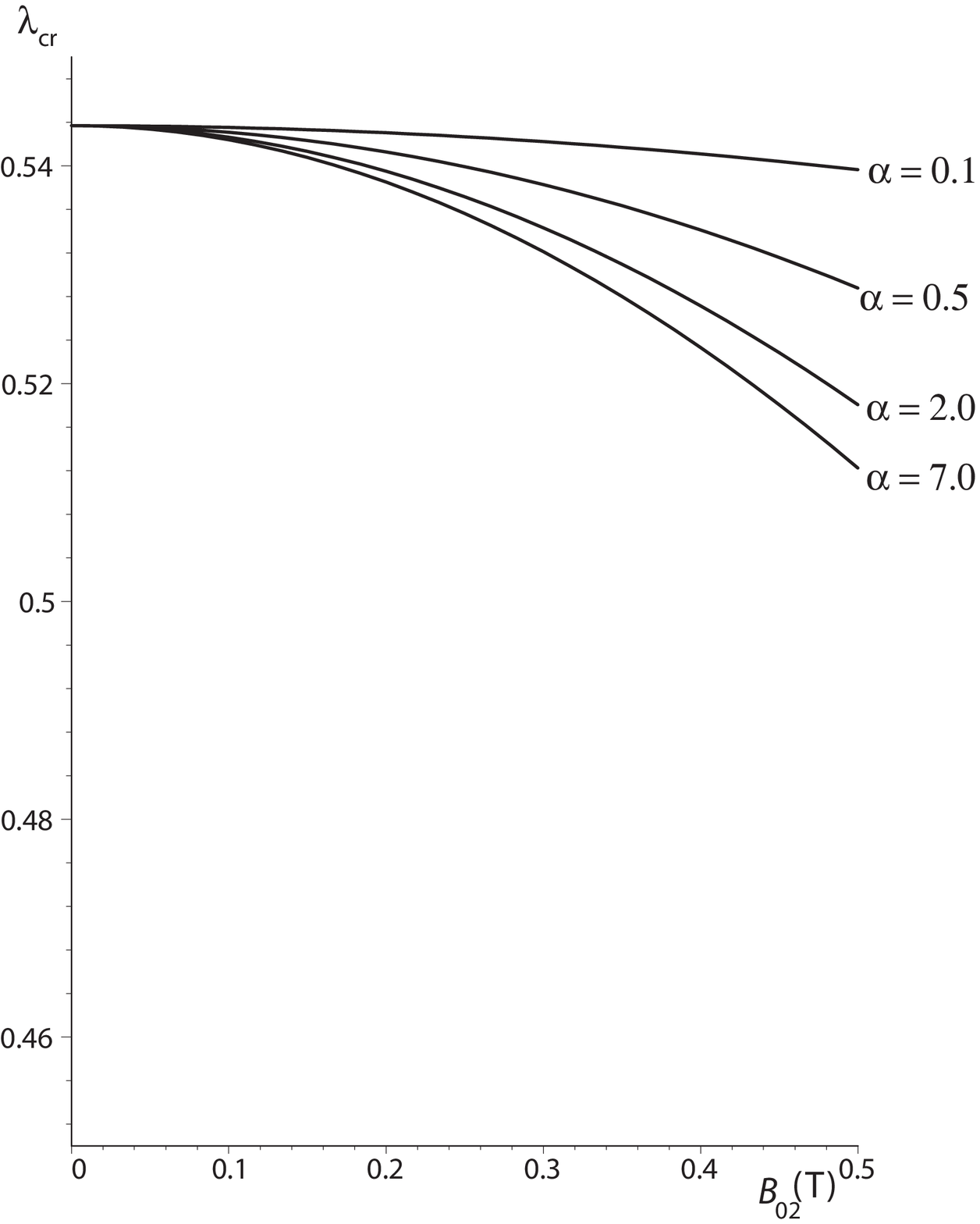,
width=.45\textwidth, height=.5\textwidth}}
 \caption{Dependence of the critical stretch $\lambda_\text{cr} < 1$ for instability in
 compression for a magnetoelastic Mooney-Rivlin solid in plane strain
 with the dimensional measure $B_{l2}$ of the magnetic field,
 for several values of the magnetoelastic coupling parameters
 $\alpha$ and $\beta$.}
\end{figure}


\appendix
\section{Derivatives of the invariants with respect to $\vec{F}$ and $\vec{B}_l$} \label{Appendix1}

We derive the expressions for the first derivatives of the six
invariants with respect to $\vec{F}$,
\begin{align}
&  \dfrac{\partial I_1}{\partial F_{i\alpha}} =2 F_{i \alpha},
\qquad \dfrac{\partial I_2}{\partial F_{i\alpha}} = 2 (c_{\gamma
\gamma} F_{i \alpha}-c_{\alpha \gamma} F_{i \gamma}),
\notag \\[0.1cm]
&  \dfrac{\partial I_3}{\partial F_{i\alpha}} =2 I_3 F_{\alpha
i}^{-1}, \qquad \dfrac{\partial I_4}{\partial
F_{i\alpha}}=0,\qquad\dfrac{\partial I_5}{\partial F_{i\alpha}} =2
B_{l\alpha} (F_{i \gamma} B_{l\gamma}),
\notag \\[0.1cm]
&  \dfrac{\partial I_6}{\partial F_{i\alpha}} =2(F_{i\gamma}
B_{l\gamma} c_{\alpha \beta} B_{l\beta}
       + F_{i \gamma} c_{\gamma \beta} B_{l\beta}
       B_{l\alpha}),
\end{align}
and with respect to $\vec{B}_l$,
\begin{align}
&  \dfrac{\partial I_1}{\partial B_{l\alpha}} = 0, &&
\dfrac{\partial I_2}{\partial B_{l\alpha}}=0, && \dfrac{\partial
I_3}{\partial B_{l\alpha}}=0,
\notag \\[0.1cm]
&  \dfrac{\partial I_4}{\partial B_{l\alpha}}=2 B_{l\alpha}, &&
\dfrac{\partial I_5}{\partial B_{l\alpha}}=2 c_{\alpha \beta}
B_{l\beta}, && \dfrac{\partial I_6}{\partial B_{l\alpha}}=2
c_{\alpha \gamma} c_{\gamma \beta}
    B_{l\beta}.
\end{align}

The second derivatives of the invariants are computed as follows:
first, the second derivatives with respect to $\vec{F}$,
\begin{align}
\dfrac{\partial^2 I_1}{\partial F_{i\alpha} \partial F_{j
\beta}}&=2 \delta_{ij} \delta_{\alpha \beta},
 \notag \\[0.1cm]
\dfrac{\partial^2 I_2}{\partial F_{i\alpha} \partial F_{j
\beta}}&=2(2 F_{i\alpha}F_{j\beta} -F_{i\beta}F_{j\alpha}+
c_{\gamma \gamma} \delta_{ij} \delta_{\alpha \beta} -  b_{ij}
\delta_{\alpha \beta} -  c_{\alpha \beta} \delta_{ij}),
 \notag \\[0.1cm]
\dfrac{\partial^2 I_3}{\partial F_{i\alpha} \partial F_{j
\beta}}&=4 I_3 F^{-1}_{\alpha i} F^{-1}_{\beta j} - 2 I_3
F^{-1}_{\alpha j} F^{-1}_{\beta i},
 \notag \\[0.1cm]
\dfrac{\partial^2 I_4}{\partial F_{i\alpha} \partial F_{j
\beta}}&=0,
 \notag \\[0.1cm]
\dfrac{\partial^2 I_5}{\partial F_{i\alpha} \partial F_{j
\beta}}&=2 \delta_{ij} B_{l\alpha} B_{l\beta},
 \notag \\[0.1cm]
\dfrac{\partial^2 I_6}{\partial F_{i\alpha} \partial F_{j
\beta}}&= 2 [\delta_{ij} ( c_{\alpha\gamma} B_{l\gamma} B_{l\beta}
+ c_{\beta\gamma} B_{l\gamma}B_{l\alpha}) +  \delta_{\alpha \beta}
F_{i\gamma}B_{l\gamma}F_{j\delta}B_{l\delta}
\notag \\[0.1cm]
 & \qquad \qquad \qquad
 + F_{i\gamma}B_{l\gamma}F_{j \alpha} B_{l\beta} +
 F_{j\gamma}B_{l\gamma}
 F_{i \beta} B_{l\alpha}+ b_{ij} B_{l\alpha} B_{l\beta}];
\end{align}
next, the mixed derivatives with respect to $\vec{F}$ and
$\vec{B}_l$,
\begin{align}
& \dfrac{\partial^2 I_1}{\partial F_{i \alpha} \partial
B_{l_\beta}}=0, \quad
  \dfrac{\partial^2 I_2}{\partial F_{i \alpha} \partial B_{l_\beta}}=0, \quad
  \dfrac{\partial^2 I_3}{\partial F_{i \alpha} \partial B_{l_\beta}}=0, \quad
  \dfrac{\partial^2 I_4}{\partial F_{i \alpha} \partial B_{l_\beta}}=0,
 \notag \\[0.1cm]
& \dfrac{\partial^2 I_5}{\partial F_{i \alpha} \partial
B_{l_\beta}}=
  2 \delta_{\alpha \beta} F_{i \gamma} B_{l_\gamma}+2 B_{l_\alpha} F_{i \beta},
  \notag \\[0.1cm]
& \dfrac{\partial^2 I_6}{\partial F_{i \alpha} \partial
B_{l_\beta}} = 2 F_{i \beta} c_{\alpha \gamma} B_{l_\gamma} + 2
F_{i \gamma} B_{l_\gamma} c_{\alpha \beta} + 2 F_{i \gamma}
c_{\gamma \beta} B_{l_\alpha} + 2 \delta_{\alpha \beta}
F_{i\gamma}c_{\gamma\delta}B_{l\delta};
\end{align}
finally, the second derivatives with respect to $\vec{B}_l$,
\begin{align}
&  \dfrac{\partial^2 I_1}{\partial B_{l\alpha} \partial
B_{l\beta}}=0, && \dfrac{\partial^2 I_2}{\partial B_{l\alpha}
\partial B_{l\beta}}=0, && \dfrac{\partial^2 I_3}{\partial B_{l\alpha} \partial
B_{l\beta}}=0,
\notag \\[0.1cm]
&  \dfrac{\partial^2 I_4}{\partial B_{l\alpha} \partial
B_{l\beta}}=2 \delta_{\alpha \beta}, && \dfrac{\partial^2
I_5}{\partial B_{l\alpha} \partial B_{l\beta}}= 2 c_{\alpha
\beta}, && \dfrac{\partial^2 I_6}{\partial B_{l\alpha} \partial
B_{l\beta}}=2 c_{\alpha \gamma} c_{\gamma \beta}.
\end{align}


\end{document}